\documentclass{aa}  

\usepackage{graphicx}
\usepackage{txfonts}
\usepackage{color} 
\usepackage{hyperref}
\usepackage{tabularx}
\usepackage{booktabs}
\font\fiverm=cmr5
\def\teq#1{$\, #1\,$}
\def\rns{R_{\hbox{\fiverm NS}}}

\begin{document}

   \title{Searching for quasi-periodicities in short transients: the curious case of GRB 230307A}

   \author{Daniela Huppenkothen
          \inst{1,2}
          \and
          Matthew G. Baring \inst{3}
          \and
          Mete Uzuner
          \inst{4,5}
          \and
          Ersin G\"o\u{g}\"u\c{s} \inst{4}
          \and
          Yuki Kaneko \inst{4}
          \and
          Chryssa Kouveliotou \inst{6}
          \and
          Oliver J. Roberts \inst{7}
          \and
          Lin Lin \inst{8}
          }

   \institute{Anton Pannekoek Institute for Astronomy, University of Amsterdam, Science Park 904, 1098 XH Amsterdam, The Netherlands\\
        \email{d.huppenkothen@uva.nl}
         \and
         SRON Netherlands Institute for Space Research, Niels Bohrweg 4, 2333 CA Leiden, The Netherlands
         \and
         Rice University,
         Department of Physics and Astronomy - MS 108,  6100 Main Street, Houston, Texas, 77251-1892, USA
         \and Sabanc\i~University,Faculty of Engineering and Natural Sciences, Tuzla, \.Istanbul, 34956, T\"urkiye
         \and
         Cardiff University, School of Physics and Astronomy,
         Queen’s Buildings, The Parade 5, Cardiff CF24 3AA, UK
         \and
         Department of Physics, The George Washington University, 725 21st Street NW, Washington, 20052 DC, USA
         \and
         Science and Technology Institute, Universities and Space Research Association, 320 Sparkman Drive, Huntsville, 35805, AL, USA
         \and
         Department of Astronomy, Beijing Normal University, Beijing, 100875, China
             }

   \date{}
 
  \abstract
   {Gamma-ray bursts (GRBs) are the most powerful explosions in the Universe; their energy release reaches us from the end of the re-ionization era, making them invaluable cosmological probes. GRB\,230307A is the second brightest GRB ever observed in the 56 years of observations since the discovery of the phenomenon in 1967. Follow-up observations of the event at longer wavelengths revealed a lanthanide-rich kilonova with long-lasting X-ray emission immediately following the prompt gamma-rays. Moreover, the gamma-ray light curve of GRB\,230307A exhibits high amplitude variability, especially within the first 15 s.}
   {We performed a timing analysis of the prompt emission of GRB\,230307A collected with \textit{INTEGRAL}'s SPectrometer of \textit{INTEGRAL} AntiCoincidence Shield (SPI-ACS) and \textit{Fermi}'s Gamma-Ray Burst Monitor (GBM).}
   {We use Fourier analysis, wavelets and Gaussian Processes to search for periodic and quasi-periodic oscillations (QPOs) in the prompt gamma-ray emission of GRB 230307A. We critically assess all three methods in terms of their robustness for detections of Quasi-Periodic Oscillations (QPOs) in fast transients such as GRBs.}
   {Our analyses revealed QPOs at a frequency of $\sim$1.2 Hz (0.82 s period) near the burst's peak emission phase, consistent across instruments and detection methods. We also identify a second, less significant QPO at $\sim$2.9 Hz (0.34 s) nearly simultaneously. We hypothesize that the two QPOs originate from the transition epoch at the end of the jet acceleration phase. These QPOs represent plasma circulation periods in vorticity about the jet axis carried outwards to the prompt radiation zone at much larger radii. They are sampled by colliding structures (e.g., shocks) in the spinning jet, possibly marking the evolution of plasma rotation during the final stages of the progenitor neutron star coalescence event.}
   {}


   \maketitle

\section{Introduction}

The prompt emission of Gamma-Ray Bursts (GRBs) in X-rays and $\gamma$-rays shows a complex time evolution \cite[see e.g.~][for an overview]{peer2015}. Unlike supernovae with their well-ordered rise and decay patterns, GRB prompt light curves are extremely variable on short timescales, with a broad diversity in their temporal structure among the burst population.  The origin of this variability is still largely unknown.  There have been numerous studies of the minimum variability timescales $t_{\rm var,min}$ by various groups, including rapid rise times. In studies with the Compton Gamma Ray Observatory's (GGRO) Burst And Transient Source Experiment (BATSE) experiment and with \textit{Fermi}'s Gamma-Ray Burst Monitor (GBM),  $t_{\rm var,min}$ values around or shorter than a millisecond are evident in the 100 keV-1 MeV band \citep{Walker-2000-ApJ,MacLachlan-2013-MNRAS}. At the higher energy, $>100$ MeV band accessed by \textit{Fermi}'s Large Area Telescope (LAT), the variability can still be on timescales of less than a tenth of a second \citep{Aldrich-2024-MNRAS}. Such $t_{\rm var,min}$ timescales yield estimates of the physical size of the GRB emission regions typically on the scale of light minutes (i.e., $\sim 1$ AU), and have provided clues toward the presence of a structured, weakly magnetized jet \citep{camisasca2023}.

Searches for periodic or quasi-periodic signals in GRB prompt emission are particularly interesting, because the specific timescales implied by a quasi-periodic oscillation (QPO) imposes strong constraints on the possible underlying emission mechanism and can potentially constrain the central engine. Some short GRBs have recently been identified as extragalactic Giant Flares from magnetars, strongly magnetized neutron stars known for their extraordinary bursting behaviour in X-rays and $\gamma$-rays \citep{roberts2021,trigg2024}. Galactic Giant Flares have exhibited both periodic oscillations associated with the neutron star's rotation period, and QPOs associated with torsional vibrations of the magnetar \citep{Strohmayer-2005-ApJ,Israel-2005-ApJ}. 
Kilohertz QPOs were seen in GRB 200415 (Castro-Tirado et al. 2021), indicating the nature of  a magnetar Giant Flare; these QPOs were subsequently identified as possible overtones of crustal oscillations using numerical simulations (Soltani et al. 2023).

Various QPOs have been claimed and contested in the literature in both long and short GRBs, using a range of different statistical methods and tests, sometimes with contradictory results. Of particular note is the contested detection in GRB 090709A, where a QPO was reported with a period of around 8\,s in \textit{Swift}, \textit{Konus}, \textit{Suzaku} and \textit{INTEGRAL} data \citep{markwardt2009,golenetskii2009,goetz2009}, but where subsequently \citet{cenko2010} showed that removing the overall trend from the light curve can lead to significant false positive detections, and similarly a reanalysis by \citet{deluca2010} also excluded a significant oscillation. Many studies have focused on the very large catalogue of GRBs observed with the \textit{CGRO}/BATSE instrument. A systematic search for QPOs in 44 bright  short GRBs observed with \textit{Fermi}/GBM, \textit{Swift}/BAT and \textit{CGRO}/BATSE provided no detections \citep{dichiara2013}, and similarly, \citet{liu2024} searched for QPOs in 532 short GRBs observed with BATSE using Fourier-based techniques and reported no compelling detections. Conversely, \citet{tarnopolski2021} report on the detection of 34 QPOs in the prompt emission of GRBs observed with BATSE using an approach based on wavelets, and \cite{Chirenti-2023-Nature} present two kilohertz QPOs in short bursts GRB 910711 and GRB 931101B, also observed with BATSE. A systematic search for QPOs in precursors of both short and long GRBs have yielded no significant detections so far \citep{xiao2022}, except for a candidate in the \textit{Swift}/BAT and \textit{Fermi}/GBM observations of the precursor of the kilonova-associated long-duration GRB 211211A \citep{xiao2024}. \citet{chirenti2024} reported a 19.5 Hz oscillation in GRB 211211A, thought to be the result of a merger between a neutron star and a black hole. Overall, the view of QPOs in GRBs is somewhat unclear: different methods come to different conclusions on the same data sets, and some detections have been contested on statistical grounds, leaving behind an uncertainty about the robustness and reliability of the existing detections.

Searches for QPOs in GRBs most often employ one of two approaches (or sometimes both): Fourier analysis and wavelets. Fourier periodograms have a long history across astronomy \citep[e.g.~][]{vanderklis1989} and other areas of science for the detection and characterisation of periodic and quasi-periodic signals, and are thus well-tested and well-understood. As we will detail more below, however, Fourier periodograms struggle with non-stationary signals, i.e.~time series whose statistical properties change as a function of time. In such cases, wavelets are often considered the prime alternative for their ability to characterize time-variable signals \citep[e.g.~][]{foster1996}. Recently, Gaussian Processes have been presented as a possible alternative in QPO searches in short transients \citep{huebner2023}, and have been explored specifically for GRBs by \citet{song2024}. 

On 2023 March 7 at 15:44:06.67 UTC, multiple space-based $\gamma$-ray telescopes detected GRB\,230307A \citep[see e.g.,][]{GCN33405, GCN33407} originating from the direction of the Magellanic Bridge. The event's T$_{90}$ duration \citep{Kouveliotou1993} was 41.5 s \citep{Sun2023}; its gamma-ray fluence (10$-$1000\,keV) reached the level of 3 $\times$ 10$^{-3}$ erg cm$^{-2}$ \citep{Sun2023}, making it the second-brightest GRB ever observed. Further, a lanthanide-rich kilonova coincident with the position and time of the GRB was identified \citep{Yang2023,Levan2023}, indicating that this event was due to a merger of two compact objects, rather than the core collapse of an evolved star. \citet{Sun2023} performed a comprehensive broadband spectral analysis and identified emission likely related to a magnetar\footnote{Magnetars are extremely magnetized, isolated neutron stars. Their spin periods are clustered in a narrow range from 0.5--12 s. \citep[For a review on magnetars, see][]{KaspiBel}} central engine, suggesting that a QPO search might reveal the magnetar's spin period. 

In this paper, we present a thorough QPO search\footnote{Code related to this paper is available at \url{https://github.com/dhuppenkothen/GRB230307A_QPOSearch}} of the prompt emission of GRB\,230307A observed with both \textit{Fermi}/GBM and \textit{INTEGRAL}'s SPectrometer of \textit{INTEGRAL} AntiCoincidence Shield (SPI-ACS). We apply all three major QPO search methods: Fourier analysis, wavelets and Gaussian Processes. We report on a candidate detection with a period of ~0.82s present across instruments and detection methods, and a less significant candidate with a period of $\sim 0.34\,\mathrm{s}$. Beyond this specific search for QPOs in this specific event, we also critically evaluate the assumptions that each method makes of the data, and whether these assumptions are fulfilled by GRBs more generally and this event in particular. As we show, all of the currently available statistical methods for QPO searches in GRBs and other short transients make strong assumptions that are not supported by the data: as a result, the differing assumptions made by each method can lead to strong disparities in the significance of a candidate signal and no clear, robust conclusion about the presence of that signal. 

In Section \ref{sec:data}, we present the data used in this study. Section \ref{sec:fourier} presents our results using Fourier-based methods, Section \ref{sec:wavelets} the analysis of the same data using wavelet transforms, and Section \ref{sec:gps} a QPO search using Gaussian Processes. For all three techniques, we critically evaluate the robustness of the method and the significance of the candidate signals. Finally, Section \ref{sec:discussion} connects these candidates with potential physical mechanisms, and considers the ability of current methods to robustly detect and characterise quasi-periodic signals in short transients.

\section{Data}
\label{sec:data}

We analyzed data collected with two $\gamma$-ray instruments: the Anti-Coincidence Shield (ACS) of the spectrometer (SPI) on board the INTErnational Gamma-Ray Astrophysics Laboratory (INTEGRAL) and the \textit{Fermi}/Gamma-ray Burst Monitor (GBM). SPI-ACS provides GRB lightcurves above 80\,keV with 50\,ms time resolution \citep{vonkienlin2003}\footnote{data downloaded from \url{https://www.isdc.unige.ch/integral/science/grb\#ACS}}. GBM detectors provide Time-Tagged Events (TTE) with a minimum readout capability of 2\,$\mu$s in the 8$-$700\,keV, 200$-$10000\,keV bands for the GBM/NaI and GBM/BGO detectors, respectively. We analysed data of two \textit{Fermi}/GBM detectors (NaI\,10 and BGO\,1).  All remaining GBM detectors either had detector zenith to source angles exceeding 60$^{\rm o}$ or were blocked by other parts of the spacecraft \citep{GCN33551}, thus not suitable for analysis.
\begin{figure*}[h!]
    \centering
    \includegraphics[width=1.0\textwidth]{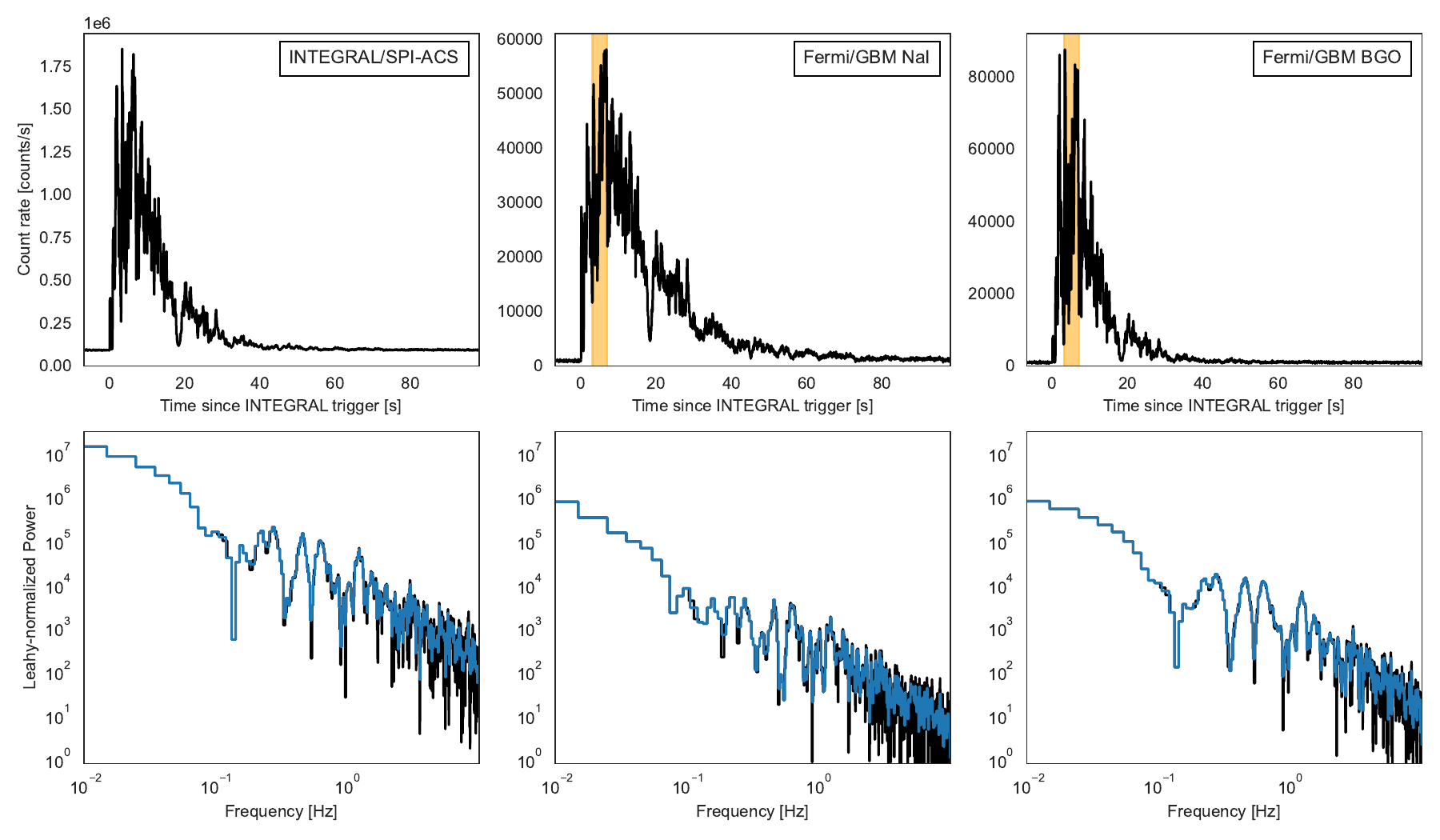} 
    \caption{Top: Light curves of GRB\,230307A in 50 ms temporal resolution as seen with SPI-ACS (left panel), and the brightest GBM NaI and BGO detectors; NaI 10 (middle panel) and BGO 1 (right panel). The time is in seconds since the trigger time. The orange area within the lower plots marks the time interval of 2.5 to 7.5 s, for which the GBM team issued a warning for possible data problems. Bottom: Fourier periodograms corresponding to the GRB light curves on the top. We show both the unbinned periodogram (black) and the log-binned periodogram (blue). Note that for the \textit{Fermi}/GBM data, these do include the segment for which a warning was issued. All three periodograms contain strong variability above the instrumental noise limit at all frequencies considered here, and show peaks on top of the broadband variability present across all frequencies in the periodogram.}
    \label{fig:lightcurves}
\end{figure*}

For the GBM detectors, we used TTE data binned to 50\,ms time resolution. When binning the TTE data we corrected for the detector deadtime \citep[$\tau=2.6\,\mu$s;][]{Meegan2009} using the nonparalysable formula:
\begin{equation}
    n = \frac{m}{1-m\tau},
\end{equation}
where $n$ is the corrected count rate and $m$ is the recorded count rate \citep{Knoll}. Deadtime introduces complex features into the periodogram, but is unlikely to strongly affect our timing analysis on very long timescales compared to the deadtime interval. At long timescales, nonparalysable deadtime such as that observed from \textit{Fermi}/GBM effectively reduces the amplitude of any present signal \citep[see, e.g.~][]{bachetti2015,huppenkothen2022}.

\textit{INTEGRAL}/SPI-ACS saturates above $\sim 1.8\times10^{6}$ counts/s \citep[e.g.~][]{savchenko2024}. We checked the \textit{INTEGRAL} light curve for bins that reach or exceed that saturation limit, and found only two bins. To check the effect of these bins, we perform the analysis in Section \ref{sec:fourier} both with the unaltered data and with a light curve where the two bins above the saturation limit are replaced by the average of its neighbouring bins. This allows us to perform a periodogram analysis, which requires evenly sampled data without gaps. Similarly, we perform the wavelet analysis in Section \ref{sec:wavelets} both with the unaltered data, and with a light curve that has the two time bins in question removed. We find no appreciable differences between the results of the analysis, and thus in this paper solely report the results derived using the original light curve.

Due to the high count rates of the event, GBM TTE data suffered data loss between 2.5\,s and 7.5\,s after the trigger time (Bad Time Interval, BTI; \citealt{GCN33551}), where the  data packets are lost due to the bandwidth limit between the instrument and the spacecraft. 
With this caveat in mind, we proceed by analysing the full light curves for both \textit{Fermi}/GBM and \textit{INTEGRAL}/SPI-ACS, as well as light curves where the BTI has been excised (where the methodological approaches described below allow for light curves with gaps). We converted all photon arrival times to the solar system barycenter to align the observations and remove any potential effects due to spacecraft orbital motion.

To determine the background level for each instrument, we fit a linear function to data segments in the preburst ($-$25 to $-$5 s) and postburst (100 to 120 s) intervals using the entire energy range. Then, we subtracted the background level from each corresponding time series to obtain the background-free burst data. While the T$_{90}$ duration of this burst is 41.5 s \citep{Sun2023}, we perform the search in the 0$-$60 s interval to capture any variability in the remainder of its gamma-ray emission phase. 

Figure \ref{fig:lightcurves} (left panels) shows the light curve of the kilonova-associated GRB, which exhibits high degree of flux variability, motivating us to search for possible periodic or quasi-periodic modulations in the lightcurves of this event.
\begin{figure*}
    \centering
    \includegraphics[width=\textwidth]{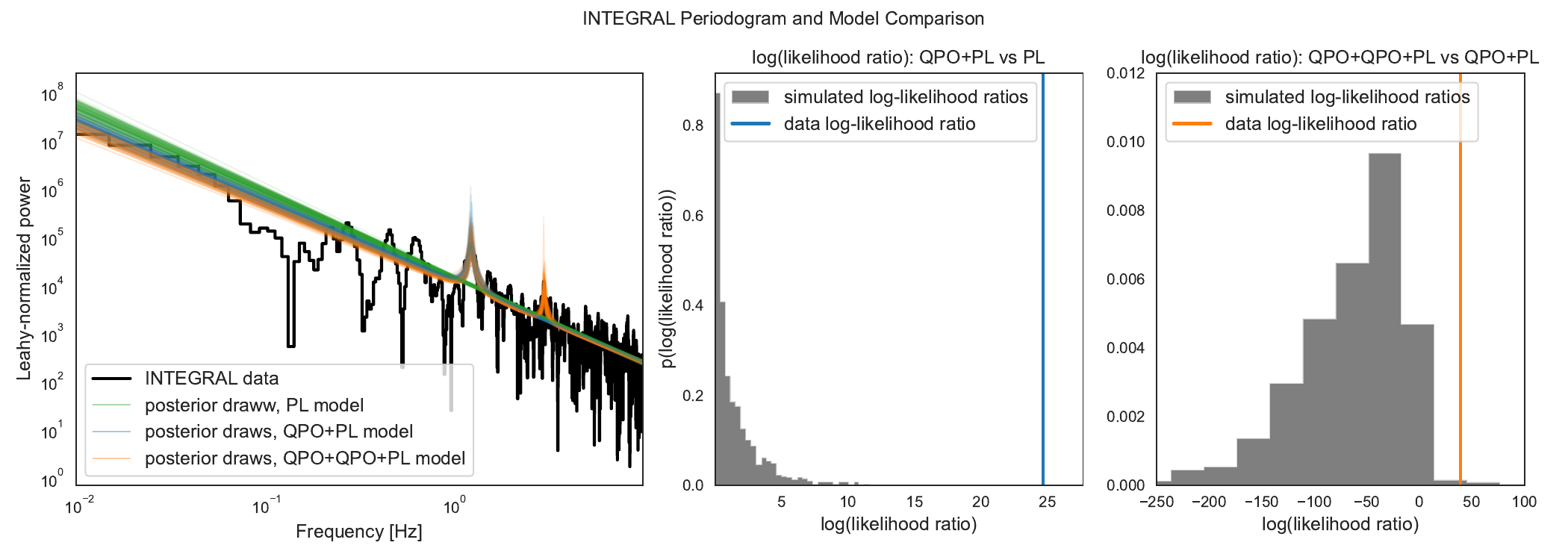}
    \caption{Left: Fourier periodogram of the \textit{INTEGRAL} data with posterior draws from the three models compared via likelihood ratio tests: in green, the power law model; in blue, a power law model with a Lorentzian component for a single QPO; in orange, a model comprising a power law and two Lorentzians. Middle: distribution of the likelihood ratios from 1000 simulated periodograms: the likelihood ratio for the observed periodogram is a clear outlier. Right: same as middle panel, but for the model with two QPOs. Again, the observed likelihood ratio is a clear outlier compared with the null hypothesis (a single QPO).}
    \label{fig:integral_psd_lrt}
\end{figure*}
\section{Fourier-based methods \& analysis}
\label{sec:fourier}

The most common standard method to search for periodic and quasi-periodic signals in astronomical data uses the Fourier transform. An introduction into the formalism and statistical background can be found in \citet{vanderklis1989} and \citet{bachetti2023}. We produced periodograms for the full GRB light curve for each instrument, normalized using the formalism by \citet{leahy83}, and show the periodograms in Figure \ref{fig:lightcurves} (bottom row of panels). The periodograms show a high amount of excess power above the Poisson noise level across the entire frequency range we probed, and in particular a set of broad, prominent features in all three instruments peaking near 0.45, 0.6, and 1.2 Hz.

\subsection{Search for (quasi-)periodic oscillations}
We followed the approach introduced by \citet{vaughan2010}, adapted to transients by \citet{huppenkothen2013}, and implemented in the \texttt{stingray} Python library \citep{stingray1, stingray2} to search for periodicities and narrow QPOs. First, we fit a model for the broadband timing variability at frequencies below $10\,\mathrm{Hz}$. We choose between a simple powerlaw and a broken powerlaw by applying a Likelihood Ratio Test (LRT). The LRT is calibrated using simulations from the simpler model, drawn from the posterior for the parameters of the power law model. We chose uniform priors between $0$ and $5$ for the power law indices, and wide, log-uniform priors for the normalization of the model, and for the break frequency. We sampled the powerlaw model using Markov Chain Monte Carlo (MCMC) as implemented in the package \texttt{emcee} \citep{foremanmackey2013}, and checked visually and through computation of the autocorrelation time whether the chains had converged. We then simulated 1000 periodograms from the posterior, and computed the LRT for each, in order to compute a tail probability ($p$-value) for the observed LRT. For all three datasets, $p > 0.05$, suggesting that the null hypothesis (a power-law model) cannot be rejected. 

We subsequently simulated another 1000 periodograms from the posterior for this model, performed a maximum likelihood fit of the power-law model for each, and computed the highest outlier in each periodogram. The null hypothesis that the maximum power in the observed periodogram can be explained by intrinsic, non-periodic variability can be rejected if the tail probability for the observation based on the sample of simulated highest outliers is small. For all three instruments, this null hypothesis cannot be rejected ($p > 0.05$).

Outlier detection methods are most powerful when the putative signal is concentrated in a single frequency bin, as is the case for a very narrow QPO or a strictly periodic signal. As shown in Figure \ref{fig:lightcurves}, the peaks visible below 1.5 Hz are distributed across multiple frequencies. To search for these signals, we perform another likelihood ratio test, but comparing the power-law model for the broadband noise with one that also includes a Lorentzian component to model a putative QPO. As for the power law, we used a wide, log-uniform prior for the Lorentzian's normalization and a log-uniform prior across the entire frequency range of the spectrum for the Lorentzian centroid frequency. We parametrized the Lorentzian's width in terms of the quality factor $q = \nu_c/\Delta\nu$ where the quality factor is defined as the ratio between the centroid frequency and the Full Width at Half Maximum (FWHM), and applied a uniform prior between $2$ and $100$. We compare that LRTs with those calculated for periodograms simulated from the null hypothesis (only the power law). 

For all three instruments, we confidently reject the null hypothesis with $p < 0.001$, with a putative QPO detection at $1.2 \, \mathrm{Hz}$ (see Figure \ref{fig:integral_psd_lrt} for details for the \textit{INTEGRAL} data; additional figures for the NaI and BGO data are in Appendix \ref{sec:appendix_psds}, Figures \ref{fig:fermi_nai_periodogram} and \ref{fig:fermi_bgo_periodogram}). This is somewhat unsurprising: the null hypothesis involves a stochastic process where powers in neighbouring bins are statistically independent. It is obvious from the periodograms that this is not the case here: for the peaked structures in the periodogram, where neighbouring bins are clearly correlated in some way, it would make sense that the LRT highly favours a model with a Lorentzian. Adding a second Lorentzian component to the model and comparing that model with the powerlaw and single Lorentzian also yields a significant rejection of the latter model in the \textit{INTEGRAL} data ($p = 0.006$), but not in the \textit{Fermi}/GBM NaI ($ p = 0.033$) or \textit{Fermi}/GBM BGO detectors ($p = 0.199$). 
We found including additional Lorentzians beyond two challenging in practice, even when exploring a wide range of starting parameters. Often, these components would optimize to local minima and very broad, flat features as part of the broadband variability. 

Overall, Fourier analysis suggests that there exist at least one, and possibly two QPOs in the data. We sampled the posterior for both QPO components along with the powerlaw model for all three instruments in order to obtain credible intervals on the QPO properties. For \textit{INTEGRAL}, we find a centroid frequency of the first QPO components of $\nu_1 = 1.217^{+0.0281}_{-0.027}$ and a full width at half maximum (FWHM) $\Delta \nu_1 = 0.094_{-0.059}^{+0.093}$, corresponding to a quality factor of $\nu_1/\Delta \nu_1 = 12.9$. Parameter estimates for the BGO data are consistent, while the results for the NaI data suggest a larger FWHM, $\Delta \nu_1 = 0.32^{+0.21}_{0.16}$ (though we note the substantially larger errors on the NaI result). For the higher-frequency QPO in the \textit{INTEGRAL} data, we find a centroid frequency of $\nu_2 = 2.987^{+0.083}_{-0.035}$ and fairly large credible intervals for the QPO width, $\Delta \nu_2 = 0.120^{+0.133}_{-0.095}$. 

\begin{figure}
    \centering
    \includegraphics[width=\linewidth]{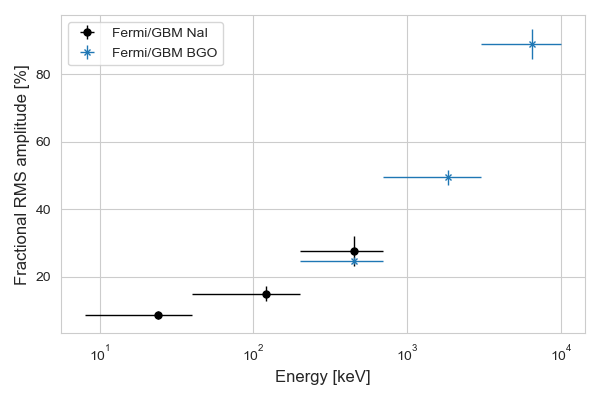}
    \caption{Fractional rms amplitude as a function of photon energy for the two \textit{Fermi}/GBM detectors.}
    \label{fig:fracrms}
\end{figure}

To check for any energy dependence of the QPOs, we generated \textit{Fermi} light curves in different energy ranges: $8-40\,\mathrm{keV}$, $40-200\,\mathrm{keV}$ and $200-700\,\mathrm{keV}$ for the NaI detectors, and $200-700\,\mathrm{keV}$, $700-3000\,\mathrm{keV}$ and $3000-10000\,\mathrm{keV}$ for the BGO detectors. The shared energy band between the two GBM detectors serves as a useful cross-check for consistency. Since the SPI-ACS detector does not record the energy of incoming photons, we cannot perform a similar analysis on the \textit{Integral} data \citep{vonkienlin2003}. For the \textit{Fermi}/GBM light curves, we find that the fractional rms amplitude of the QPO at $1.2\,\mathrm{Hz}$ is strongly energy-dependent (Figure \ref{fig:fracrms}, increasing from $\sim 16\%$ to $\sim 88\%$ from the lowest to the highest-energy band. This behaviour persists across the detectors, and the fractional rms amplitude for the overlapping band is consistent within statistical uncertainties. Because of the relatively lower significance, and the wide credible intervals on the QPO width for the second QPO in the \textit{Fermi}/GBM data, we do not perform a similar analysis for the potential QPO candidate at 2.9 Hz.

The increase of the RMS amplitude with energy band might well be expected for a wide array of physical emission mechanisms.  The time variability of the flux at different energies will naturally couple to spectral variability: the spectral shape will likely not be preserved as the flux goes up and down, and at energies where the spectrum is steeper, fluxes are likely to vary more.  This effect is most pronounced if the break energy varies.  When this happens, the fractional changes in fluxes are likely to be greater at a fixed photon energy above the break than at a fixed energy below the break where the spectrum is flatter.  For this burst, Band model fits \citep[see e.g., Table~2 of ][]{Dichiara-2023-ApJL} indicate that the break energy $E_{\rm br}\sim 635-970$ keV and the spectral indices $\alpha$ (below $E_{\rm br}$) and $\beta$ (above) all vary substantially over the time interval pertinent to the QPO analysis. Thus, fluctuations in $E_{\rm br}$ are likely to lead to a higher RMS amplitude for QPOs above this energy than below it.

\begin{figure*}
    \centering
    \includegraphics[width=\textwidth]{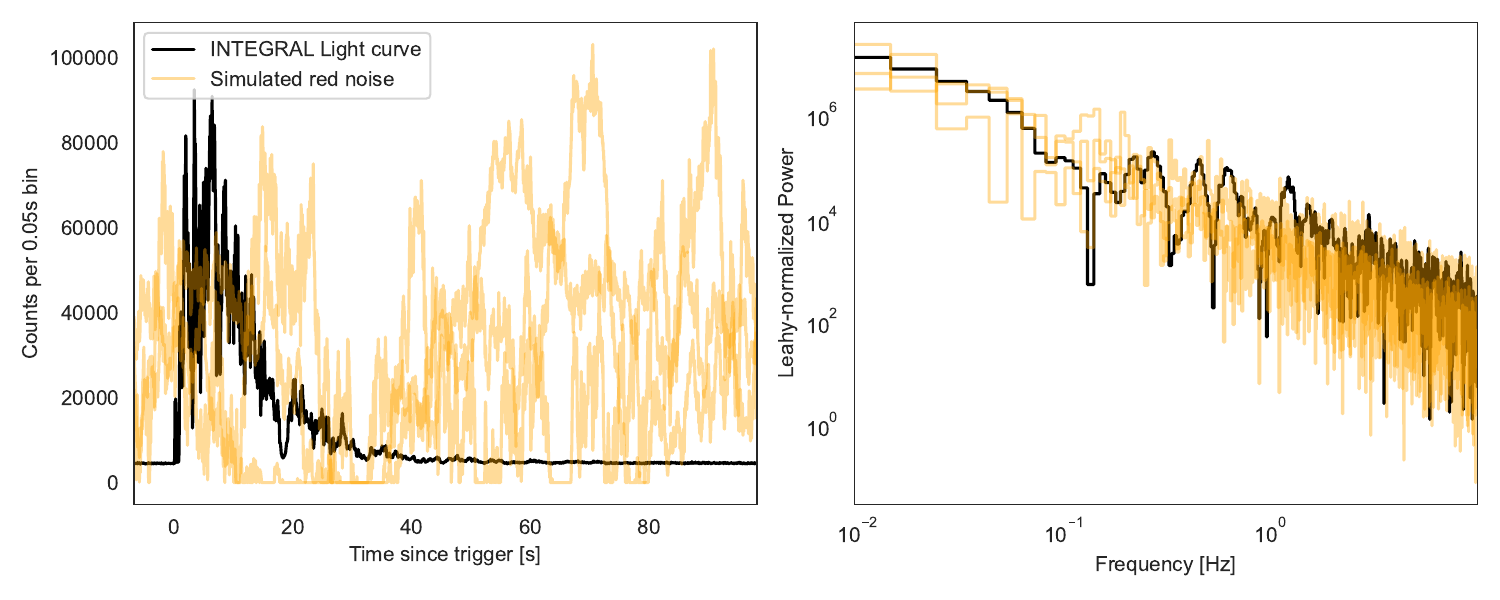}
    \caption{Left: \textit{INTEGRAL} light curve of GRB230307A (black), with three random light curves generated from the stochastic process used in the QPO detection methods outlined in this section (orange). The GRB has a well-defined beginning and an end, in between which there exists rapidly changing variability. The simulated light curves also contain variability at a high amplitude, but the overall process does not change throughout the light curve. This is expected for a stationary stochastic process. Right: periodogram of the \textit{INTEGRAL} data (black) and of the simulated light curves (orange). While the periodogram of the GRB exhibits peaks formed by excess power in correlated neighbouring frequency bins, the periodograms of the stochastic process contain--by design and construction--powers that are statistically independent.}
    \label{fig:psd_process}
\end{figure*}
\subsection{Assumptions and limitations}

The Fourier-based methods used in this section are well-tested and well-understood in the context of periodicity and QPO searches in astronomical light curves. As described in \citet{vanderklis1989} and \citet{bachetti2023}, we understand the statistical properties of periodograms of stochastic processes observed with X-ray and gamma-ray instruments very well. However, these methods make specific assumptions that are a challenge in QPO detections in fast transients such as GRBs. In particular, one assumption that underpins the analysis above is that of weak stationarity. Weak stationarity (or wide-sense stationarity) assumes that the mean of the process and the autocovariance of the process do not change as a function of time. This is generally true in the context of X-ray binaries and Active Galactic Nuclei over the timescales of interest for QPO searches, where many of these methods are routinely applied successfully. However, it is obviously not true for short transients: by definition, they have a beginning and an end, and as with the case of this GRB, may show very significant changes in the variability as a function of time. As shown in \citet{huebner2022}, a key consequence of this non-stationarity is that powers in neighbouring bins are no longer statistically independent. Any method that relies on that independence, as those above do, will tend to overestimate the significance of any candidate signal. 

The mismatch between the model above and the data considered here is easily illustrated by generating periodograms from the posterior for the parameters of the model with two QPOs, and simulating light curves from those periodograms using the method in \citet{timmer1995}. We show the results in Figure \ref{fig:psd_process}. The mismatch between the GRB and the process assumed for both the overall variability and the possible QPO is immediately obvious: while a stationary stochastic process can still display large-amplitude variability, the properties of this variability do not change as a function of time, whereas for the GRB they do. 

One possible solution to the above problem could be to de-trend the light curve. This would entail fitting a model to the overall shape of the burst, subtracting the best-fit model, and then continuing with the Fourier analysis assuming the residuals follow a stationary stochastic process as assumed by the statistical methods above. However, this approach also comes with a range of challenges. First and foremost, it assumes that the observed light curve can be neatly separated into an overall burst shape and a stationary stochastic process. This may be the case, but the presence of variability from the stochastic process will significantly bias the fit to the overall burst, and thus also introduce biases into the residuals that again lead to a departure from the statistical assumptions of the method. Additionally, removing an overall trend generally removes power from the smallest frequencies in the periodogram. As \citet{cenko2010} showed, doing so may lead to artifacts in the periodogram of the residuals that can mimic a QPO, and thus lead to spurious detections, a result recently confirmed by \citet{song2024}. Finally, in this particular case, it is unlikely that detrending will solve the underlying problem. In all three instruments, the amplitude of the variability appears to be time-dependent, that is, the amplitude of the variability in the first ten seconds or so is significantly larger than twenty seconds after the burst. Thus, removing an overall trend will not yield residuals that can be modelled with a stationary stochastic process. We thus do not proceed with detrending. 

\section{Wavelets}
\label{sec:wavelets}

A common method applied to short bursts and other astronomical transients are wavelets \citep[e.g.~][]{hurley1998,morris2010,golkhou2014}. The wavelet transform is in some ways a generalization of the Fourier transform to a wider range of basis functions beyond sines and cosines. Crucially, most common wavelet functions are localized in both time and frequency. As a result, they do not require stationarity and are often used, for example, to detect time-dependent periodic and quasi-periodic signals.

\subsection{Search for (quasi-)periodicities}

\begin{figure*}
    \centering
    \includegraphics[width=\textwidth]{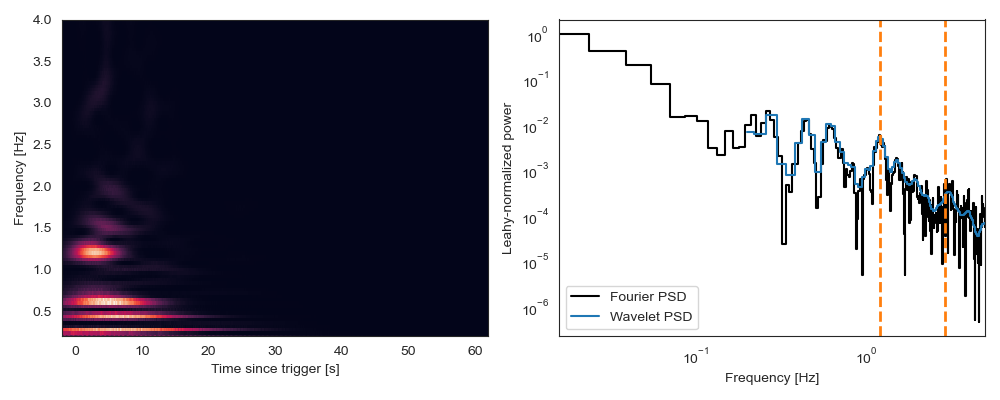}
    \caption{Left: 2D wavelet transform (spectrogram) of the \textit{INTEGRAL} observation of GRB230307A. The transform shows transient power at low frequencies in the first $\sim$20 seconds or so of the burst, where the variability is particularly strong. It also shows a short, transient signal between $0-10\,\mathrm{s}$ at 1.2 Hz, similarly to what was identified in the Fourier periodogram. The candidate signal at 0.32 Hz identified in the Fourier periodogram is less apparent here. Right: Fourier periodogram (black) and wavelet periodogram (blue) with the candidate signals found in the Fourier analysis noted as orange dashed lines. The Fourier and wavelet periodograms largely match.}
    \label{fig:wavelets}
\end{figure*}

We perform a wavelet transform with the Python library \texttt{PyWavelets} \citep{Lee2019} using a complex Morlet wavelet, defined as

\begin{equation}
\psi(t) = \frac{1}{\sqrt{\pi \beta}}\exp{\left(-\frac{t^2}{\beta}\right)}\exp{(2\pi i \nu_c t)} \; ,
\end{equation}

\noindent where $\beta$ is the bandwidth and $\nu_c$ is the centre frequency. The bandwidth $\beta$ describes the spread of power in the frequency domain (corresponding to a time decay in the time domain). The centre frequency $\nu_c$ should be chosen near the frequencies of interest to be explored using the wavelet transform. The complex Morlet equation above describes a complex exponential windowed by a zero-centred Gaussian, with a width set by $\sqrt{\beta/2}$. We choose a centre frequency of $\nu_c = 1.5 \,\mathrm{Hz}$, broadly in the range where we expect to see signals based on the periodogram, and a bandwidth of $\beta = 10$, corresponding to around 20 rotational cycles. We show the two-dimensional wavelet transform and the one-dimensional wavelet periodogram for the \textit{INTEGRAL} data in Figure \ref{fig:wavelets}. 

The two-dimensional wavelet transform (or spectrogram) shows strong power at the lowest frequencies, and also clearly shows the candidate signal at 1.2 Hz. The candidate signal at 2.9 Hz, however, is less visible, though there are some faint structures at higher frequencies. All power is concentrated early in the light curve, suggesting that any signal is transient and only present in the first $\sim$10-20 seconds. The wavelet periodogram (Figure \ref{fig:wavelets}, right panel) is very consistent with the Fourier periodogram, as expected for this kind of wavelet. 

\begin{figure*}
    \centering
    \includegraphics[width=\textwidth]{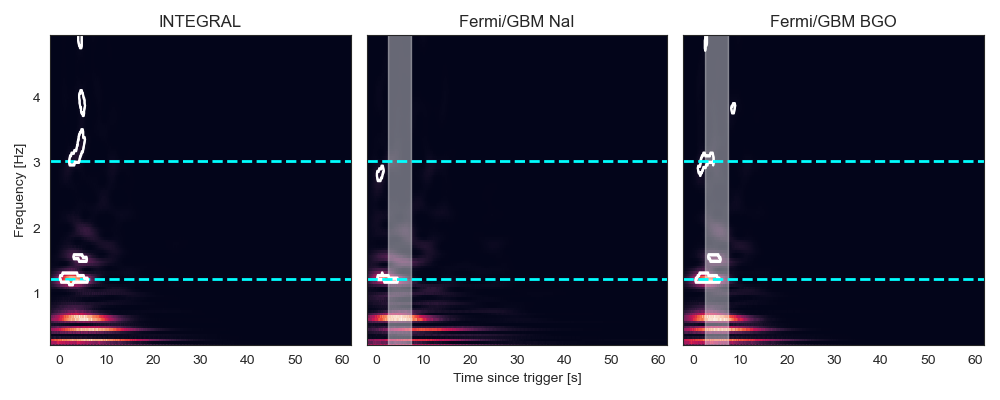}
    \caption{Wavelet spectrograms for GRB 230307A for all three instruments. In all three, we find significant power at the lowest frequencies, as well as power at $1.2\,\mathrm{Hz}$. In white contours, we overplot the 99.99\% percentiles. In the \textit{Fermi}/GBM data, we mark the segment flagged by the \textit{Fermi} team for potential data issues as the shaded region.}
    \label{fig:wavelet_results}
\end{figure*}

Estimating significance in wavelet spectrograms is generally a challenging task. The standard approach in the literature 
\citep[e.g.][]{ghosh2023} takes a variability model (often a power law-like red noise model or equivalent) and generates simulated light curves from this model, and significance is then determined using per-bin outlier tests comparing the observed wavelet power in a bin to the distribution of simulated wavelet power in the same bin. We follow this approach and simulate 20,000 light curves from the posterior for the power law model generated in Section \ref{sec:fourier} using the approach in \citet{timmer1995}, and subsequently produce wavelet spectrograms for each simulated light curve in the same way as for the real data. We then compute 99.99\% percentiles for each time-frequency bin, and identify those bins where the observed wavelet power exceeds the 99.99\% percentile derived from the simulations. We perform this analysis independently for each data set (i.e.~INTEGRAL, \textit{Fermi}/GBM NaI and BGO detectors), using their individual posterior distributions for the power law model. The results are presented in Figure \ref{fig:wavelet_results}. This approach clearly identifies regions in the spectrogram where the observed wavelet power is an outlier compared to the simulations with $p < 0.0001$. In all three instruments, the candidate signal at 1.2 Hz is clearly detected, and present only in the first $\sim\,10$\,s or so of the burst. In \textit{Fermi}/GBM, this overlaps with much of the interval flagged by the \textit{Fermi}/GBM team for potential data issues (BTI). However, we also note the similarity of the period, width and temporal extent of the signal in all three instruments, which may suggest that these data issues do not significantly affect the detection of the QPO. We find a weaker area of significance at the candidate signal around 2.9 Hz identified in the Fourier analysis. This, too, is highly localized to a small segment of the overall burst within the first $\sim\,5$\,s after the trigger, though we also note that the signal evolves temporally to a higher frequency with time in all three instruments. There are additional small regions where bins exceed the significance threshold, but we find that these are not consistent in time or frequency across the three instrument, and thus consider them likely spurious candidates.

The results presented in Figure \ref{fig:wavelet_results} constitute single-trial $p$-values, and are thus not corrected for the number of trials. Estimating the number of trials for a wavelet spectrogram is challenging because for all but very specific choices of frequency and time resolution, neighbouring bins in the spectrogram are not statistically independent. In addition, a trial-corrected $p$-value calculated using the approach above would require millions of simulations given that we would need to correct for 76680 frequency bins (making the conservative but incorrect assumption of statistical independence), which is computationally prohibitively expensive. The fact that there are patches of significance rather than individual pixels does not necessarily constitute additional evidence for the existence of a QPO: because neighbouring bins are not statistically independent, observing patches of high significance can be a result of that lack of independence, rather than evidence for the existence of a real signal. 

We also perform the most conservative possible estimation of significance, and compare the single bin with the highest wavelet power to the distribution of the highest wavelet power found across all simulations, and find that using this metric, none of the candidate QPOs are significant ($p = 0.1$ for both \textit{INTEGRAL} and \textit{Fermi}/GBM NaI data, respectively; $p = 0.075$ for the \textit{Fermi}/GBM BGO data). However, note that this is likely too conservative, because it does assume statistical independence of spectrogram bins.

\begin{figure}
    \centering
    \includegraphics[width=0.45\textwidth]{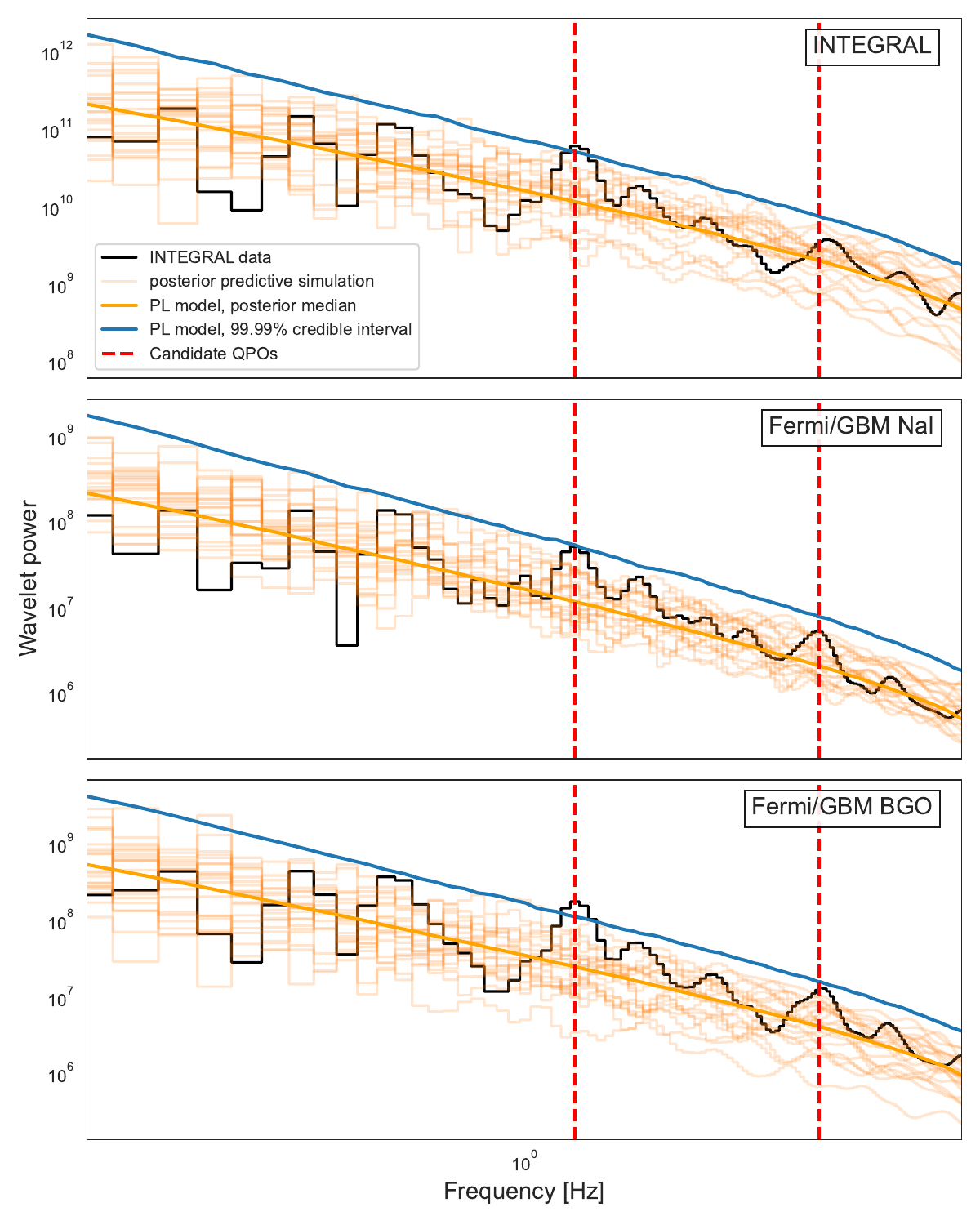}
    \caption{In black, we show the wavelet periodogram for the data (top: \textit{INTEGRAL}; middle: \textit{Fermi}/GBM NaI; bottom: \textit{Fermi}/GBM BGO). The wavelet periodogram corresponds to the $2D$ wavelet spectrogram integrated over the time axis. In orange, the posterior mean derived from $1000$ simulated wavelet periodograms, along with 10 posterior draws from the power-law stochastic model sampled in Section \ref{sec:fourier}. In blue, we show the $99.99\%$ single-trial detection limit. The candidate QPO at 1.2 Hz exceeds that limit for all three instruments, while the candidate QPO at 2.9 Hz exceeds this limit only for the \textit{Fermi}/GBM BGO data. Note, however, that integrating over the time axis will necessarily yield lower significances given the short-lived nature of both candidate signals in the wavelet spectrogram.}
    \label{fig:wpsd_results}
\end{figure}

Finally, we use the simulations to compute single-trial significances in the wavelet periodogram integrated over time (Figure \ref{fig:wpsd_results}). This analysis is similar to the Fourier periodogram in that integrating over the time dimension means we lose the advantage of time-dependent modelling of the wavelet spectrogram. Calculating single-trial significances at a significance threshold of $p < 10^{-4}$ corresponds to a trial-corrected significance threshold of $p < 0.012$, or approximately $3\sigma$. We find that in agreement with the results from the Fourier analysis, the candidate QPO at 1.2 Hz is above this $3\sigma$ threshold in all three instruments, whereas the 2.9 Hz signal only reaches $3\sigma$ in the \textit{Fermi}/GBM BGO data. 

\subsection{Assumptions and limitations}

A key limitation of the wavelet-based method is the challenging determination of significance and the number of trial correction in the presence of correlated frequency- and time-bins. In the analysis above, we have stated both single-trial significances as well as results for the most conservative possible assumption (that all bins are statistically independent), which will vastly decrease sensitivity of the QPO detection. The truth likely lies somewhere in the middle. Given the persistence of the signal at 1.2 Hz in the wavelet periodogram and its consistency with the Fourier analysis, we consider this to be a strong candidate for a QPO. The putative signal at 2.9 Hz is a little less clear: it, too, is present in all three wavelet spectrograms, but with less consistency than the 1.2 Hz signal. Additional power in the wavelet periodogram is inconsistent between the three instruments, and we consider these likely to be statistical artifacts. We note that a fraction of the 1.2 Hz signal falls into the segment flagged as subject to data quality issues in \textit{Fermi}/GBM. However, given the consistency of the signal in \textit{Fermi}/GBM and \textit{INTEGRAL}, and that a part of the signal is present before the bad time interval, we suggest that the relevant data issues have not significantly impacted the detection of the QPO. 

There is, however, a somewhat more fundamental open challenge with the analysis above. While wavelet transforms are better suited to the detection of transient signals, and especially of transient periodic signals, we still must compare them to a model parametrizing the null hypothesis. Here, we followed the standard analysis for QPOs with wavelets in the literature and used the same stochastic power-law-shaped process as defined in Section \ref{sec:fourier}. This means that irrespective of the transform applied to the data (and simulations), the same caveats nevertheless apply: a stationary stochastic process is not a good representation of the underlying data (see also Figure \ref{fig:psd_process}), which in turn will affect the trustworthiness of the significances derived using simulations generated from this process. 

This is also the reason why we do not combine $p$-values for the three instruments for any of the analysis methods considered in this paper: doing so is only permissible if the assumption of independence holds. Here, this assumption holds only for instrumental noise, which will be generated independently in each detector, but not for stochastic variability in the GRB itself, which is produced at the source. To derive more reliable significances requires a more realistic, non-stationary model for the data, which we will define in the next section.

\section{Gaussian Processes}
\label{sec:gps}

We search for QPOs in the time domain using the method developed in \citet{huebner2023} based on Gaussian Processes (GPs; e.g., \citealt{williams2006gaussian}; for an introduction to GPs in astronomy, see \citealt{aigrain2022}). In short, the method directly models the light curve as a combination of a non-stationary trend function parametrizing the overall shape of the burst and combines this with a stochastic process to model variability on top of this trend function. Through Bayesian model comparison, different classes of models (e.g. different types of trend functions) can be compared. This model can take into account the non-stationary nature of a burst in a more principled way through the trend function, though we note that we can currently not yet implement a nonstationary QPO except in some simple forms described in \citet{huebner2023}. 
Due to the exceptionally bright nature of this GRB, the Gaussian measurement uncertainties for the data assumed in Gaussian Processes are broadly applicable.

\begin{table}[ht]
\centering
\setlength{\tabcolsep}{4pt}
%\makegapedcells
\begin{tabularx}{\columnwidth}{@{}l|X|l @{}}
\toprule    \\
\textbf{Parameter} & \textbf{Meaning} & \textbf{Distribution} \\\hline
$t_p$                      & peak of the skew-Gaussian          & $\mathcal{U}(0, 60)$ \\
$\log(A_m)$                & log-amplitude of the skew-Gaussian & $\mathcal{U}(5, 15)$ \\
$\log(\sigma_\mathrm{rise})$ & log of the rise time of the skew-Gaussian & $\mathcal{U}(-1, 3.5)$ \\
$\log(\sigma_\mathrm{fall})$ & log of the fall time of the skew-Gaussian & $\mathcal{U}(1, 4.0)$ \\ \hline
$\log(\nu_c)$              & log of the QPO centroid frequency & $\mathcal{U}(\log(1), \log(3))$ \\
$\log(A_\mathrm{QPO})$     & log-amplitude for the QPO  & $\mathcal{U}(2, 40)$ \\
$\log(C_\mathrm{QPO})$     & log-width for the QPO   & $\mathcal{U}(-20, 40)$ \\ \hline
$\log(A_\mathrm{RN})$     & log-amplitude for the red noise  & $\mathcal{U}(2, 40)$ \\
$\log(C_\mathrm{RN})$     & log-width for the red noise  & $\mathcal{U}(-20, 40)$ \\ \hline
$\log(\alpha)$  & logarithm of the AR parameters &  $\mathcal{U}(-20, 20)$ \\
$\log(\beta) $ & logarithm of the MA parameters & $\mathcal{U}(2, 20)$ \\
    \hline
    \end{tabularx}
    \caption{Prior distributions for the parameters in our model. $\mathcal{U}(a,b)$ corresponds to a uniform prior distribution between limits $(a, b)$. The first four parameters describe the skew-Gaussian mean function, followed by three parameters describing the QPO, and finally two parameters describing a Damped Random Walk stochastic process. We also include priors for the CARMA(2,1) model considered in Section \ref{sec:carma}. }
    \label{tab:priors}
\end{table}

The approach chosen here appears similar to detrending, but has the advantage that it can take into account uncertainties in the parameters of the detrending function. It can also correctly account for correlations between the parameters of that function and the variability not modelled by the trend function.
By simultaneously considering both the trend function and the variability on top of it, we can derive appropriately unbiased estimates of the trend function, and take into account uncertainties in that estimate, as well as correlations between the parameters of the trend function and the stochastic process.

We choose a skew-Gaussian function\footnote{Note that we use this here as a functional form, rather than its usual use as a statistical distribution.} as a trend function, defined as

\begin{equation}
f(t)  = A \begin{cases}
  \exp{\left(\frac{-(t-t_c)^2}{2\sigma_1^2}\right)}  & t < t_c  \\
  \exp{\left(\frac{-(t-t_c)^2}{2\sigma_2^2}\right)}  & t \geq t_c \; .
\end{cases}
\end{equation}

\noindent The skew-Gaussian function provides a flexible model for the asymmetric, approximately exponential rise and decay of the burst. We compare two stochastic processes: one model parametrizes a Damped Random Walk (DRW), the second that same DRW combined with a QPO parametrized as a stochastically driven damped harmonic oscillator. The DRW, also known as an Ornstein-Uhlenbeck Process or an autoregressive process of order 1 (AR(1)) is a fairly simple stochastic process that parametrizes the flux at time $t+1$ in terms of the flux at the previous time $t$ and a random component. In Section \ref{sec:carma}, we consider higher-order autoregressive moving-average processes as an alternative to this process.

We implement wide, uninformative priors reflecting our lack of prior knowledge in most of the model parameters (see Table \ref{tab:priors}). Many relevant parameters depend on the properties of the data: amplitudes depend on the sensitivity of the instrument, and the frequency range we can search depends on the length of the GRB. For those parameters, we set priors that reflect the ranges we can reasonably expect to see in our data. Priors on parameters for the trend function and the DRW that exist in both the model with and without QPO are the same for both. 
\begin{figure*}
    \centering
    \includegraphics[width=\textwidth]{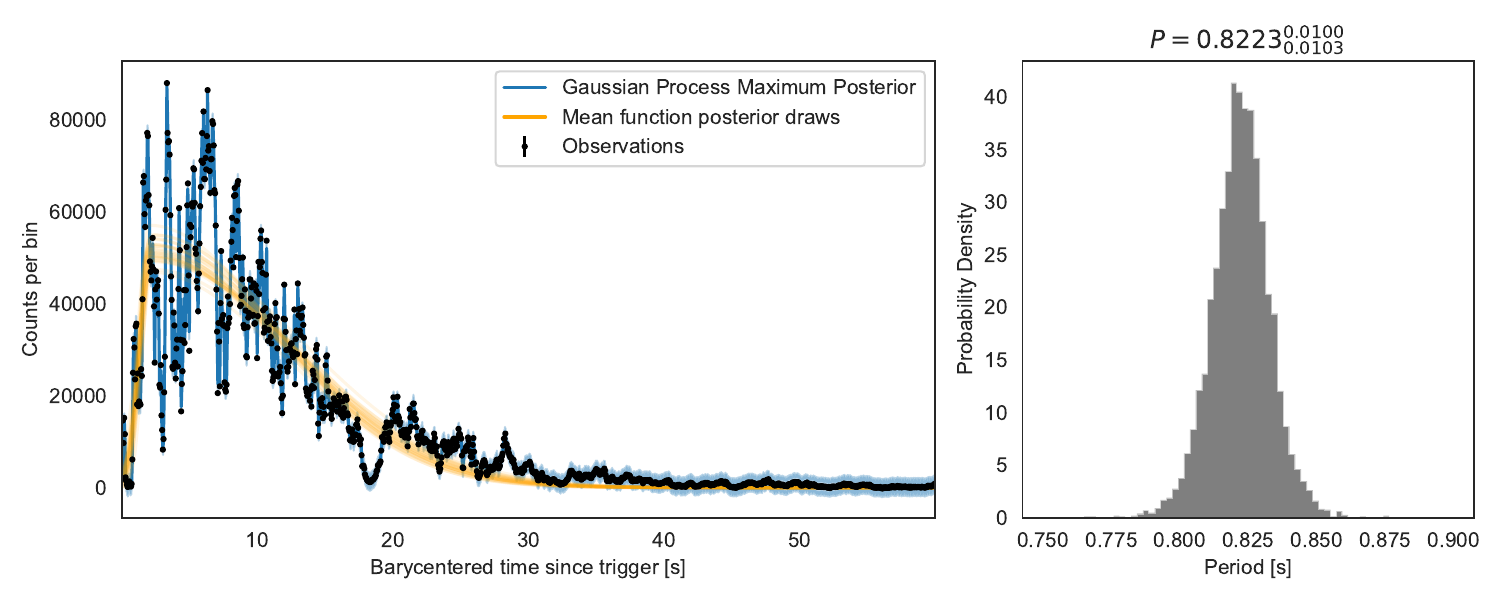}
    \caption{Left: SPI-ACS light curve (black points) in units of counts per $0.05\mathrm{s}$ bin, the predicted rates from the maximum-a-posteriori combined model consisting of the mean function, the stochastic process and a QPO (blue), and posterior draws from the mean function (orange). Right: The posterior probability density for the QPO centroid period.}
    \label{fig:integral-drw-full}
\end{figure*}

We sample each model using the Gaussian Process library \texttt{tinygp} \citep{tinygp} and sample the posterior using Nested Sampling as implemented in the Python package \texttt{jaxns} \citep{jaxns} with 2000 live points. We compare the two models using the Bayes factor $\mathcal{B}_{21}$ for the model with a QPO (model $\mathcal{M}_2$) versus a model without (model $\mathcal{M}_1$). In logarithmic form, a positive value for $\log_{10}(\mathcal{B}_{21})$ can be interpreted as evidence for the presence of a QPO component, whereas a negative Bayes factor can be interpreted as evidence against. The significance of the Bayes factor is calibrated using the common scale by \citet{kass1995}, which considers $\log(\mathcal{B}_{21}) > 2$ as decisive evidence for model $\mathcal{M}_2$. For more information on the method, including simulations to calibrate its ability to detect QPOs, see \citet{huebner2023}.

\subsection{SPI-ACS GP QPO search}
\label{sec:spi-acs}

Comparing GP models with and without QPO yields a highly significant signal centred at the frequency of 1.21$\pm$0.01 Hz with a Bayes factor of $\log_{10}(\mathcal{B}_{21})= 3.70$, indicating decisive evidence for the presence of a QPO component in the data \citep{kass1995}. The posterior distribution for the QPO's centroid frequency is narrow and unimodal, and the signal is fairly coherent, with a quality factor $\log(q) = 2.63^{+1.18}_{-0.82}$, corresponding to $q \simeq 13$, highly consistent with the estimate from the Fourier analysis. We present the corresponding posterior period distribution in the right panel of Figure \ref{fig:integral-drw-full}, which peaks at 0.82$\pm$0.01s. Overall, the posterior distributions are well-constrained for both $\mathcal{M}_1$ and $\mathcal{M}_2$ (see the corner plot in Figure \ref{fig:integral-drw-qpo-corner} in Appendix \ref{sec:appendix_psds}, generated with the Python package \texttt{corner} \citep{corner}), and uncertainties in the individual evidences used to compute the Bayes factor are small ($\sim 0.1$). The sampling results are stable across multiple runs and the Bayes factor remains significant even when changing the uninformative priors significantly. 

\begin{figure*}
    \centering
    \includegraphics[width=\textwidth]{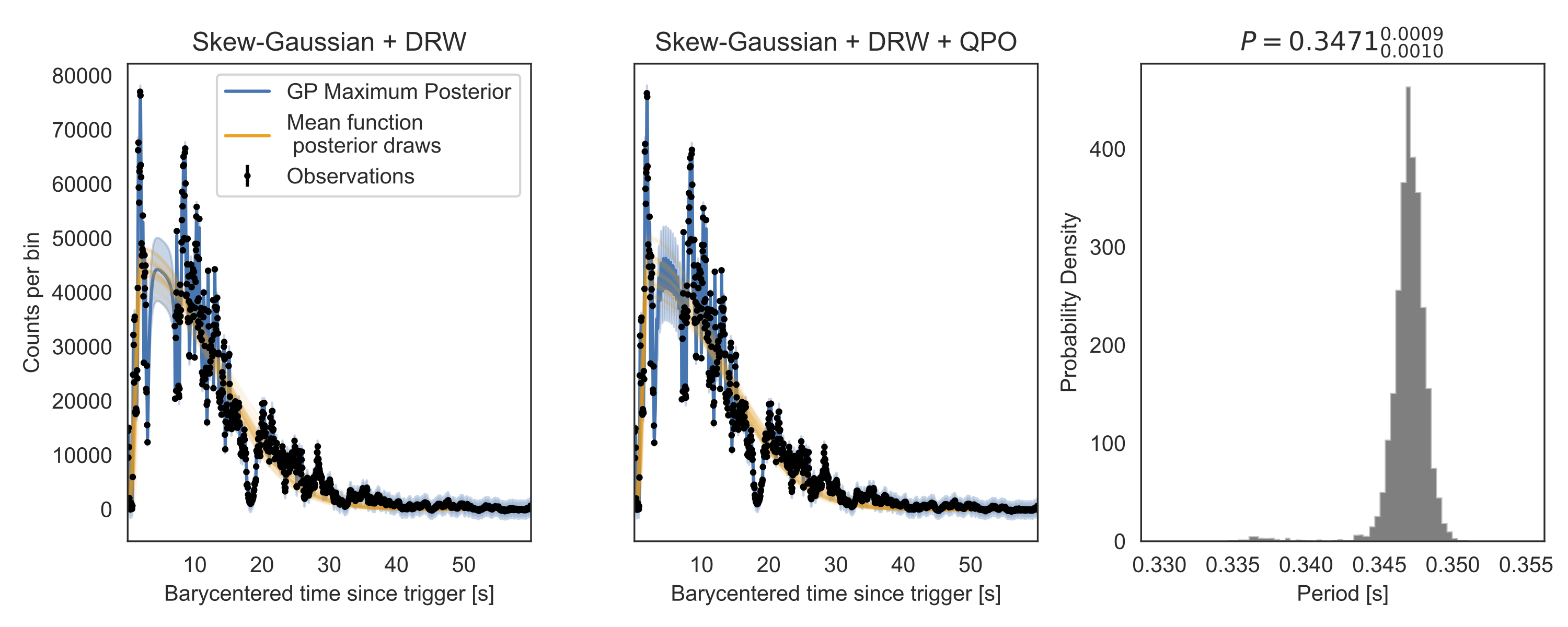}
    \caption{Re-analysis of the SPI-ACS data excising the brightest part of the GRB (i.e., BTI), which is affected by data loss in \textit{Fermi}/GBM. In the left and middle panel, we show observations in black, the maximum-a-posteriori combined model in blue, and draws from the posterior probability density for the mean function in orange. The left-most panel shows the results for $\mathcal{M}_1$ (skew-Gaussian mean function and DRW), the middle panel for $\mathcal{M}_2$ (skew-Gaussian mean function, DRW and QPO). In the right-hand panel, we show the posterior probability density for the period of the QPO, which is narrowly constrained around $P=0.347\,\mathrm{s}$.}
    \label{fig:acs_gp_no_bti}
\end{figure*}

Even though the SPI-ACS data are not affected by the same instrumental issues during the brightest intervals of the GRB as \textit{Fermi}/GBM is, we repeat the analysis with the \textit{Fermi}/GBM BTI excised, in order to directly compare results across instruments, under otherwise identical conditions (i.e.~ with the same priors and Nested Sampling settings). We find evidence of a strong QPO in this light curve as well (Figure \ref{fig:acs_gp_no_bti}), but a different frequency broadly consistent with the second candidate signal found in the Fourier and wavelet representations: $\nu_0 = 2.880\pm0.001 \,\mathrm{Hz}$, corresponding to a period of $0.347\,\mathrm{s}$. The signal's significance is $\mathcal{B}_{21} = 3.99 \pm 0.1$, indicating a decisive preference for the model with a QPO. As with the full data set, this result is stable across Nested Sampling runs and when varying the already very wide priors. The signal has a high quality factor, $\log(q) = 4.77^{+0.57}_{-0.52}$, indicating that the signal is quite coherent. We note that while this period is much shorter, it does not correspond to a harmonic of the signal detected in the full data. 

The wavelet representation suggests that much of the 1.2 Hz signal is covered by the bad time interval, and thus excised from the light curve along with that bad time interval. On the other hand, the candidate QPO at 2.9 Hz in the Fourier and wavelet representations is concentrated into a short interval before the BTI. When considering a model containing only one QPO, it makes sense that the posterior will be concentrated on the stronger of the two signals. When considering the whole light curve, this will be the 1.2 Hz QPO. However, removing the BTI, and much of the 1.2 Hz QPO with it, yields a light curve requiring a model with a QPO at 2.88 Hz.

\subsection{\textit{Fermi}/GBM GP QPO search}
\label{sec:fermi_gp_qposearch}

\begin{figure*}
    \centering
    \includegraphics[width=\textwidth]{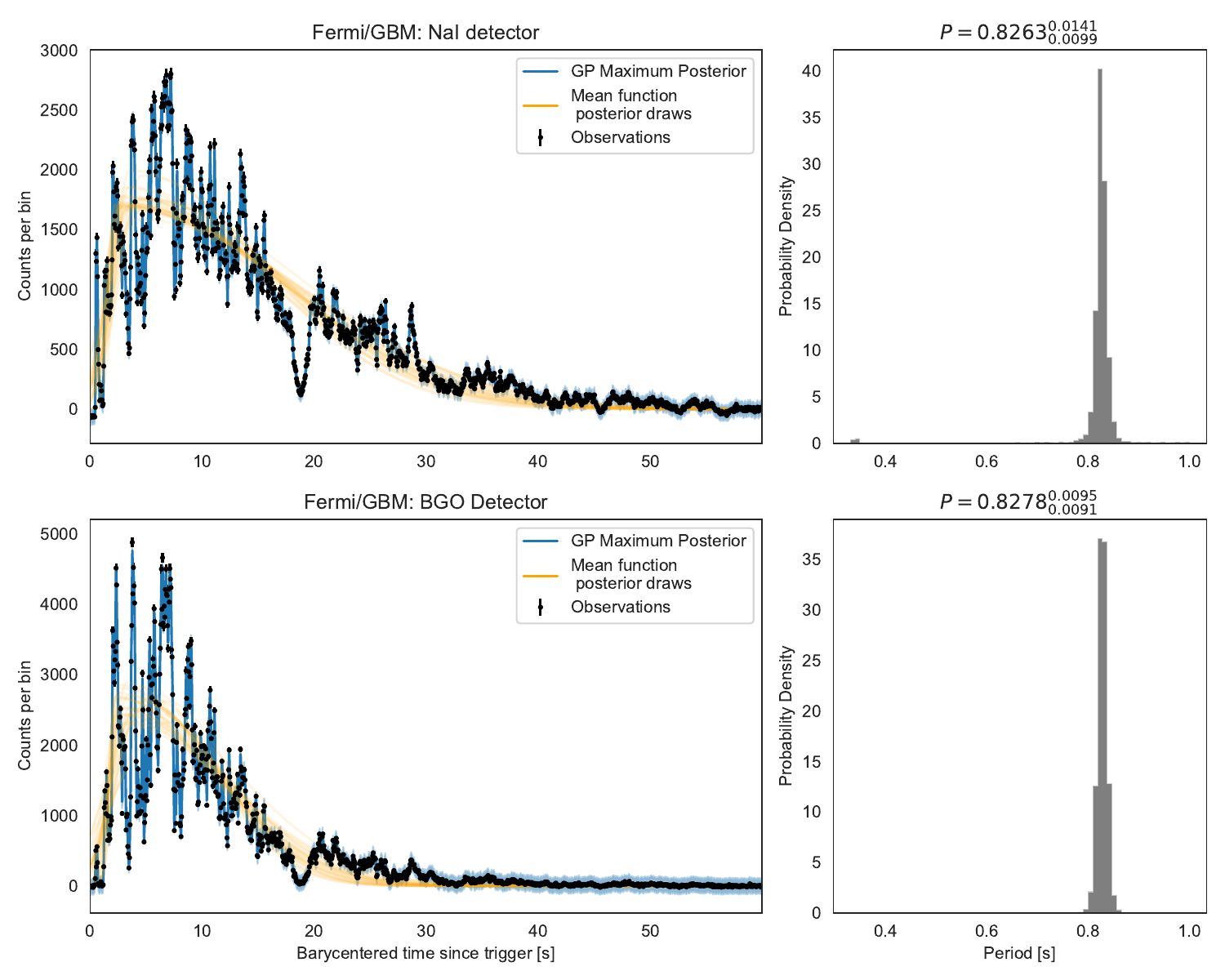}
    \caption{Left panels: light curve (black), maximum-a-posteriori model (blue) and posterior draws from the mean model (orange) for the \textit{Fermi}/GBM NaI detector (na; top) and the BGO detector (b1; bottom). Right: corresponding posterior probability densities for the period in the model including a QPO.}
    \label{fig:fermi-whole-postpred}
\end{figure*}

We performed a similar analysis with the data of both NaI (na) and BGO (b1) detectors. For each, we model the full light curve--including BTIs--with both $\mathcal{M}_1$ and $\mathcal{M}_2$, and find results that are highly consistent with the SPI-ACS data (see Figure \ref{fig:fermi-whole-postpred}). Both light curves show highly significant QPOs, with Bayes factors of $\log_{10}(\mathcal{B}_{21}) = 2.12$ (NaI detector) and $\log_{10}(\mathcal{B}_{21}) = 4.68$ (BGO detector). In both, the posterior probability density for the period is very constrained, $P_\mathrm{NaI} = 0.825^{+0.07}_{-0.02}\,\mathrm{s}$ and $P_\mathrm{BGO} = 0.826\pm 0.009\,\mathrm{s}$. The distribution for quality factors for the NaI data is somewhat broader, $\log(q) = 2.37^{+1.52}_{-3.47}$, reflecting a broader period posterior, but again we see very high coherence of the signal in the BGO detector, $\log(q) = 3.34^{+1.51}_{-1.16}$. 

The strong consistency between the results for SPI-ACS and \textit{Fermi}/GBM leads us to conclude that the data loss likely did not significantly impact the timing results. Note that the period posterior for the NaI detector shows a minor mode at $\sim 0.34$, where we found a significant signal when excluding the \textit{Fermi}/GBM BTI in the SPI-ACS data.
We also analyzed the \textit{Fermi}/GBM data with the BTI excluded in order to obtain a light curve not affected by data loss and less affected by deadtime. As with the SPI-ACS data, we find that the QPO period has shifted to $P=0.3476\pm 0.008\,\mathrm{s}$, consistent with the SPI-ACS results (Figure \ref{fig:fermi-cutout-postpred} in Appendix \ref{sec:appendix_psds}). For both datasets, we find high Bayes factors, $\log(\mathcal{B}_{21,\mathrm{NaI}}) = 5.25$ and $\log(\mathcal{B}_{21,\mathrm{BGO}}) = 5.57$, respectively, and strongly constrained, high quality factors for the QPO, $\log(q_{\mathrm{NaI}}) = 6.75^{+1.35}_{-2.73}$ and $\log(q_{\mathrm{BGO}}) = 4.69^{+0.72}_{-0.78}$.

\subsection{Detection sensitivity}

Bayes factors are notoriously sensitive to prior choices. The empirical model we employ means that we largely chose wide, uninformative priors, which in turn may affect the sensitivity to QPOs in the data. 
To calibrate the detection sensitivity for the Bayes factor, we simulated fake GRBs drawing only from the joint posterior for $\mathcal{M}_1$, i.e.~ the model without a QPO. For each draw, we generate a trend function and sample from a Gaussian Process combining that trend function with a realization of the DRW process. We add Poisson photon counting noise to this simulated GRB to generate a realistic light curve. We model this simulated light curve in the same way we do for the real SPI-ACS data, using the same models and prior assumptions. We repeat this procedure for a 100 different simulations\footnote{We limit ourselves to 100 simulations in order to keep computational requirements manageable.} drawn from the posterior, and generate a distribution of Bayes factors expected under the model without a QPO. This allows us to explore what range of Bayes factors we would expect under model $\mathcal{M}_1$.

\begin{figure}
    \centering
    \includegraphics[width=0.45\textwidth]{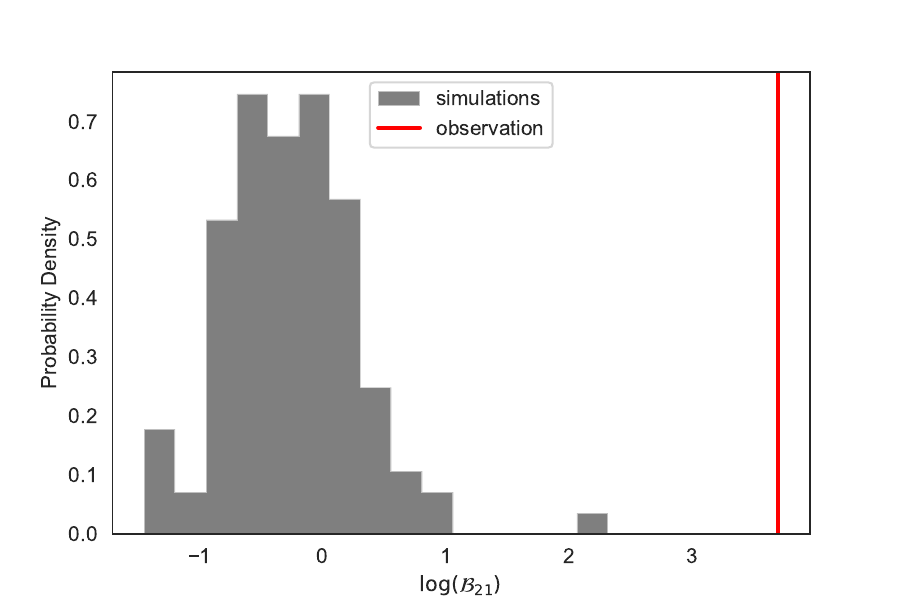}
    \caption{Distribution of the Bayes factors obtained by modeling 100 simulated GRBs based on a model consisting of a skew-Gaussian trend function and a DRW, but without a QPO (grey). The Bayes factor obtained for the SPI-ACS data is a strong outlier compared to the simulations, indicating that our analysis captures additional variability not present in the simpler model.}
    \label{fig:integral-fap}
\end{figure}

In Figure \ref{fig:integral-fap}, we present the resulting distribution; compared to the observation, we find much smaller Bayes factors for the simulated observations, indicating that we should not have observed the high Bayes factor recorded for the SPI-ACS data if the latter had been generated by a simple stochastic process and the trend function. 

\subsection{Alternative stochastic models}
\label{sec:carma}

The DRW model considered above to explain the variability in the GRB is a fairly simple model with a power spectrum constrained to Lorentzian centred at zero. While our goodness-of-fit test has shown that this model can explain the data, we nevertheless implement a somewhat more complex model: a continuous autoregressive, moving-average (CARMA) model of order $p=2$ and $q=1$. CARMA models consider both an autoregressive (AR) process and a moving average (MA) process simultaneously. Here, the AR describes the future of a system based on its current state and a random perturbation. The MA process parametrizes the time series of a system in terms of a signal and its convolution with an impulse response function. The orders $p$ for the AR process and $q$ for the MA process describe the time lags and length of the impulse response, respectively \citep{moreno2018}. The CARMA(2,1) process we implement here is capable of a wider range of power spectral shapes. We limit ourselves to this process, since higher-order CARMA processes can intrisically include QPO-like behaviour, making them not practicable as an alternative model in the context of QPO searches. We use the Python package \texttt{tinygp} for the implementation of an CARMA(2,1) process, and combine it with the same prescription for the mean function above. The priors for the parameters $\alpha$ and $\beta$ of the CARMA process are similarly wide and uninformative as for the damped random walk (see Table \ref{tab:priors}); the amplitude of the process is in this parametrization folded into the MA parameter $\beta$. 

\begin{figure*}
    \centering
    \includegraphics[width=\textwidth]{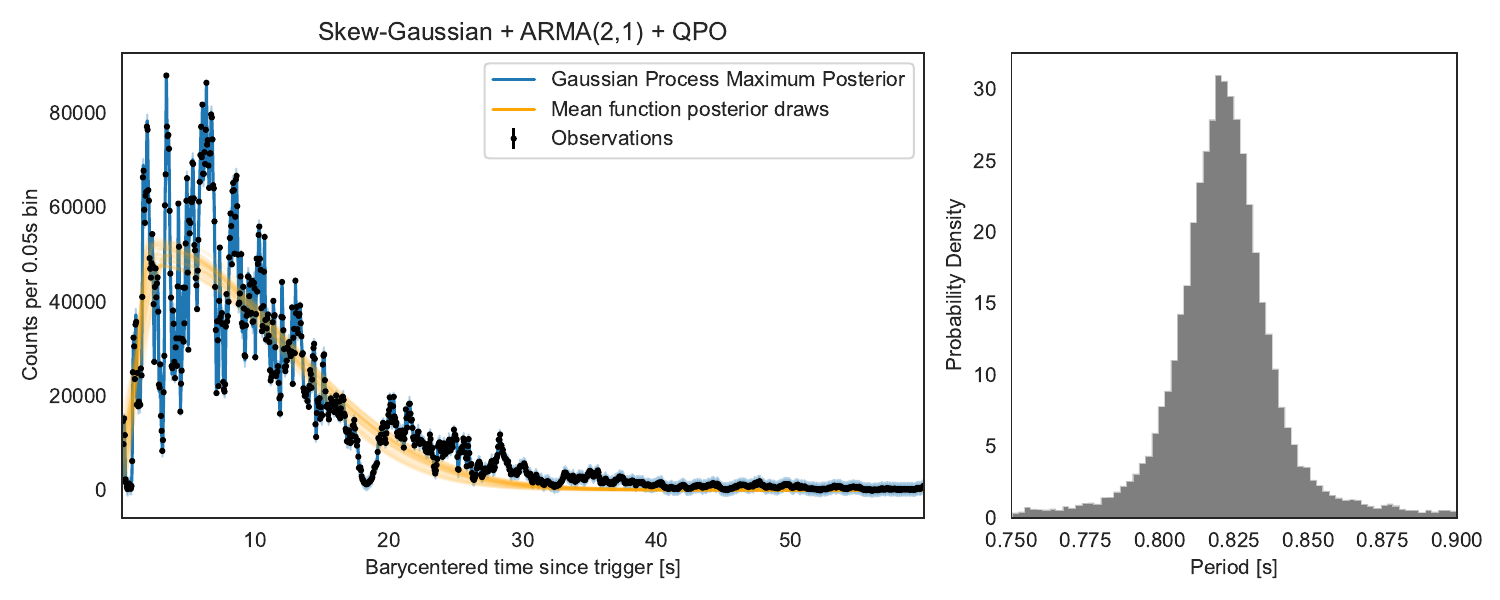}
    \caption{Left: \textit{INTEGRAL} SPI-ACS observation (black) with the maximum-a-posteriori model comprising a CARMA(2,1) process, a QPO and a skew-Gaussian mean function (blue), with posterior draws from the mean function in orange. Right: Posterior probability density of the period parameter shows a clear peak at $0.82\,\mathrm{s}$, consistent with results obtained using the simpler DRW model.}
    \label{fig:integral-carma}
\end{figure*}

We model the full light curves for all three instruments with both a mean function and a CARMA(2,1) process, and a model that additionally includes a QPO. For all three light curves, we once again find Bayes factors that strongly favour the model including a QPO ($\log_{10}(\mathcal{B}_{21}) >2$ for all three datasets). The QPO period is similarly constrained to $0.83\,\mathrm{s}$ in with this model as with the model containing a DRW (see Figure \ref{fig:integral-carma} for an example). The independence of our results of the chosen model for the underlying variability strongly suggest that the QPO is real.

\begin{figure*}
    \centering
    \includegraphics[width=\textwidth]{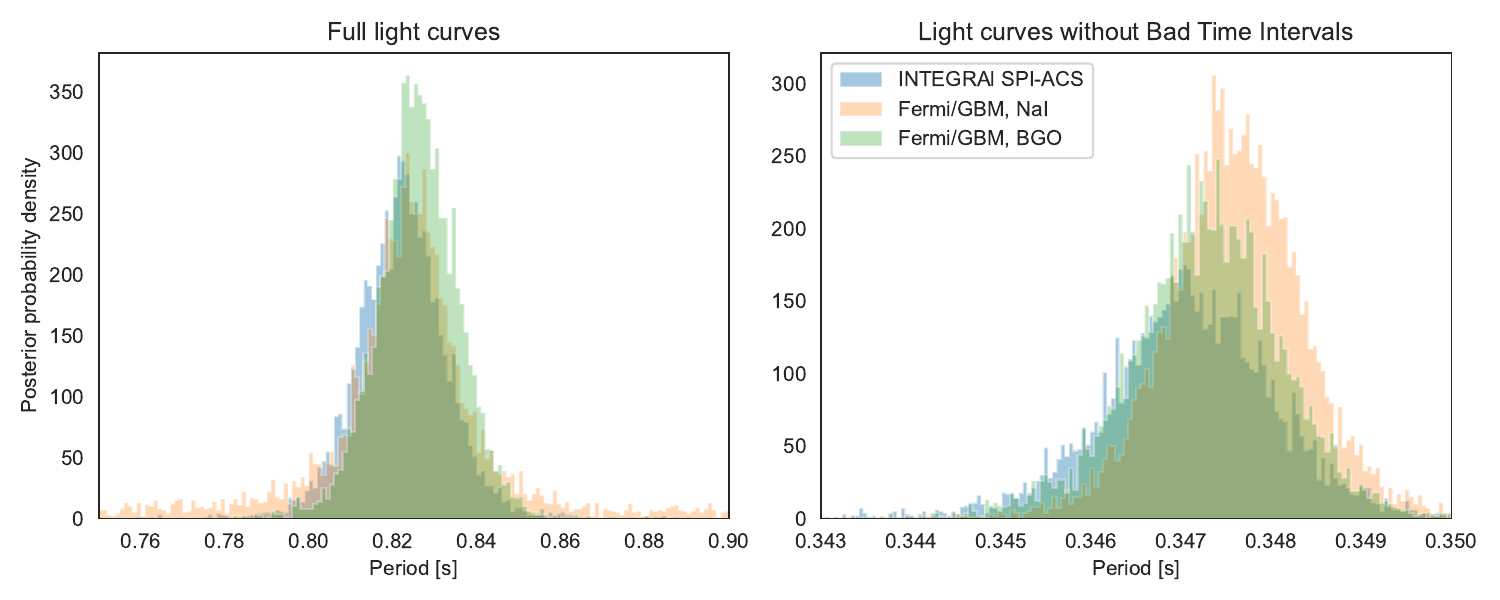}
    \caption{Period posteriors for all three instruments for the full light curves (left) and for the light curves without the \textit{Fermi}/GBM Bad Time Intervals of 2.5--7.5 s after trigger (right). The posteriors show a remarkable degree of agreement between the three instruments, at frequencies corresponding to $\sim$1.2 Hz (0.83 s period) for the full light curves and at $\sim$2.9 Hz (0.34 s) without BTI.}
    \label{fig:periods}
\end{figure*}

When we exclude the flagged segment (BTI) of the burst, we find strong signals at 2.9 Hz (0.34 s) in the data of all three instruments with similarly significant Bayes factors (see Figure \ref{fig:periods}) and results consistent with the analysis using the DRW model.

\subsection{Assumptions and limitations}

Gaussian Processes provide two key advantages over standard Fourier analysis and wavelet methods: they enable robust modelling of unevenly sampled light curves directly in the time domain, minimizing aliasing and windowing effects, and they enable to joint modelling of a stochastic process together with an overall, deterministic trend, as we do here. This comes at considerable computational cost to calculate and calibrate the Bayes factors.

A main assumption of Gaussian Process modelling is that data uncertainties are normally distributed. While not strictly true for the data considered here, GRB 230307A is bright enough for the Gaussian approximation to be justified. A second key assumption is that the GRB can be decomposed into a linear combination of a trend function parametrizing
the global rise and fall of the burst, and a stochastic process parametrizing the variability on shorter timescales. Empirically, this appears to be not a bad assumption, but we also note that none of the models implemented here--Gaussian Processes or in any of the previous methods--are physically motivated. While the DRW is a simple choice for a covariance function,  increasing the flexibility by considering a higher-order CARMA process did not substantially alter our conclusions. We note, however, that visually, the amplitude of the variability over the course of the GRB appears to change as a function of time. This is at odds with the models considered here, and implementation of a model including some form of non-linear variability component is beyond the scope of this paper. 

\section{Discussion}
\label{sec:discussion}

In this paper, we have presented a thorough analysis of the light curves of GRB 230307A taken with both \textit{INTEGRAL}/SPI-ACS and \textit{Fermi}/GBM's NaI and BGO detectors. Across multiple detection methods, instruments and wavelengths, we consistently identify a short-lived QPO at $1.2 \,\mathrm{Hz}$ (0.82s) in the first ten seconds of the GRB, when emission is at its brightest. Given that the signal is very strongly present in the SPI-ACS data, we exclude the possibility that its presence in the GBM data could be related to data issues concurrent with the signal's presence. 

A second, short-lived QPO appears to exist simultaneously in all three instruments at $2.8\,\mathrm{Hz}$ (0.34s), though not all tests return a confident detection. In the wavelet spectrogram, this signal appears to show an upward frequency trend over its lifetime. This signal is especially strongly present in QPO detection tests that excise the \textit{Fermi}/GBM BTI. Given that most QPO detection methods are designed to find a single, strongest QPO candidate, this is unsurprising: removing most of the interval containing the $1.2\,\mathrm{Hz}$ QPO naturally leaves this second candidate as the strongest signal.

All three methods used have limitations, and as we have shown, make assumptions that are not met by the light curves analysed here. In particular, the assumption of pure stochastic (red) noise made by standard analyses in Fourier and wavelet domains hampers our ability to make robust detections in counter-intuitive ways: \citet{huebner2022} showed that many approaches may overconfidently detect QPOs in this context, because non-stationary light curves break the stationarity assumption of most methods. Wavelets--while excellent at detecting non-stationary, short-lived QPOs--cannot free us from this challenge if the assumed null hypothesis remains a stationary process. As theoretical modelling of GRBs improves, physically motivated predictions for the variability expected in GRB prompt emission would dramatically improve our robustness and sensitivity to find these signals.

We quote significances for all three detections independently: combining detection probabilities across instruments or methods is generally not applicable in analyses containing substantial amounts of intrinsic source variability beyond detector noise, because multiplication of probabilities relies on strict statistical independence of the underlying tests. This is given in the case of pure instrumental noise, but much more complicated when the underlying variability is intrinsic to the source emission and thus shared across instruments (modulo some energy-dependence of the variability). However, we consistently find $3\sigma$ detections (or equivalent) in multiple methods for both QPOs, and thus conclude that both are very strong candidates. 

Physically, the signature evidence of two QPOs at frequencies in the Hz range can be explained in terms of a jet launching and evolution scenario.  We note that different values for these frequencies might be obtained if different time windows could be probed with sufficient statistical precision in $\nu$-space, and the main information is that they are not in the kHz range.  The jet birth picture will focus on a binary merger progenitor, though the scenario described can also apply to a core collapse hypernova, should kilonova associations with long-duration bursts like GRB\,230307A become more widely established.

As a binary neutron-star (NS) merger proceeds, the tidal disruption that extracts plasma from one or both stars that eventually becomes the GRB jet, only arises just before coalescence. This corresponds to an orbital semi-major axis of a few NS radii \teq{\rns} at most, and natural orbital frequencies $\Omega=2\pi\nu \sim c/(3\rns ) \sim 10^4$\,Hz for jet birth. As the merger proceeds to smaller radii $r$, the $\Omega$ value increases modestly before the plasma shedding abruptly terminates at coalescence in presumably forming a black hole.   Millisecond QPO periods from the merger proximity are likely mostly obscured by the plasma shroud, which thins out only at photospheric radii $R_{\rm ph}\sim 10^{12}\,$cm for long-duration GRBs.  Yet we note the recent report \citep{Chirenti-2023-Nature} of kHz QPOs present in two short duration GRBs from the BATSE archive.  This is an interesting result, albeit indicating a rarity of QPOs among the GRB population.

As the jet is launched, due to pressure from its surrounding disk/cocoon medium, it nominally develops a quasi-parabolic morphology \citep{Tchekhovskoy-2008-MNRAS,Komissarov-2009-MNRAS} with an extraction of significant angular momentum {\bf J} in the form of magnetic field helicity and plasma vorticity in a Poynting-flux dominated jet.  The field maintains causal connection to the rotation (i.e. $\Omega$) at the jet's \teq{r\sim \rns} base during the acceleration phase, wherein $\gamma \propto r^{1/2}$ approximately \citep{Tchekhovskoy-2008-MNRAS}.  Eventually the cocoon pressure declines and becomes insufficient to control the jet shape and dynamics, so that jet enters a coasting phase prior to its prime GRB prompt emission epoch.  While the cocoon shapes the jet and controls its 
dynamics, it acts as the boundary to an acoustic cavity of circumference $\propto r^{1/2}$ transverse to the jet axis.  Once the jet coasts, this boundary has effectively dissolved and rotational plasma fluctuations in the lateral dimension have no preferred timescale.  

For the gas, at a light cylinder radius, typically quite close to the resultant compact object, the plasma vorticity saturates as the circular speed of the gas about the jet axis nudges $c$. Thereafter, as the jet expands to larger radii $r$, its lateral (sheath) extent scales as $r^{1/2}$ while the circular speed is still $c$, so that the plasma rotation period $P$ in the observer frame increases as $P\sim 2\pi (r\, \rns)^{1/2}/c \propto r^{1/2}$. A ``freeze-out'' of the plasma $P$ occurs approximately when it enters the coasting epoch, which for fiducial jet launching models \citep[see Fig.~8 of][]{Tchekhovskoy-2008-MNRAS} is around a resultant jet $\gamma = 10^3$ at around \teq{r/\rns \sim 10^6}, leading to \teq{P\sim 0.1-1}sec. This is consistent with the QPO periods observed herein, with the rotational freeze-out arising at distances $r\sim 10^{12}\,$cm from the merger product, i.e. around the photospheric radius. More than one period may be sampled as the chaotic driver of the central engine interfaces with the cocoon sheath at the onset of the coasting and optically thin epoch.   

In a quasi-acoustic phenomenon controlled by the cavity extent lateral to the jet direction, colliding plasma structures riding the gas vorticity in the jet would subsequently sample Fourier power at this rotational period as the jet becomes optically thin.  Higher frequency Fourier power from earlier epochs would generally be muted due to high photospheric densities. After dynamic decoupling of the jet from the cocoon, plasma fluctuations have no natural acoustic driver, and so their timescales decouple from further lateral expansion of the jet.  Throughout, longitudinal fluctuations exist, but are not bounded geometrically, and so possess a chaos associated with the activity of the central driving engine, reflected in the light curves we see.

The QPOs observed thus constitute approximate images of the freeze-out plasma vorticity in a radially-structured jet acquired at the larger distances, $r_{\rm rad}\sim 10^{14}-10^{16}$cm, associated with prompt GRB radiation.  As such, they enable jet archaeology by providing a window into the cessation of the jet launch/acceleration phase that is approximately contemporaneous with the jet's exit from the photosphere. Accordingly a core goal of jet launching simulations should be the reproduction of these rotational/QPO periods at the onset of coasting. The two QPO frequencies could be an imprint of the immediate pre-merger NS-NS binary evolution, carried forth to $r_{\rm rad}$: higher frequency signals correspond to jet plasma preparation deeper in the pre-merger gravitational potential where the Keplerian period is shorter.  The observed values of $P=2\pi/\Omega$ at $\sim$0.8 and $\sim$0.35 seconds do not distinguish whether the product of the merger is a black hole, a neutron star or even a magnetar.  They could well be analogous to year-timescale periods in flux levels observed from $\gamma$-ray blazars \citep{Penil-2023-MNRAS} with their jets emanating from supermassive black holes.

\begin{acknowledgements}
We thank the anonymous reviewer for their helpful comments, Matteo Lucchini for advice on rms-energy spectra, and Volodymyr Savchenko for advice on dead time and saturation in \textit{INTEGRAL}/SPI-ACS. DH is supported by NWO's Women In Science Excel (WISE) fellowship. OJR gratefully acknowledges funding support from NASA under grant 80NSSC24M0035.
\end{acknowledgements}

\bibliographystyle{aa} 
\bibliography{aa53952-25} 

\begin{appendix}
\onecolumn

\section{Supplementary figures}\label{sec:appendix_psds}

\begin{figure*}[h!]
    \centering
    \includegraphics[width=\textwidth]{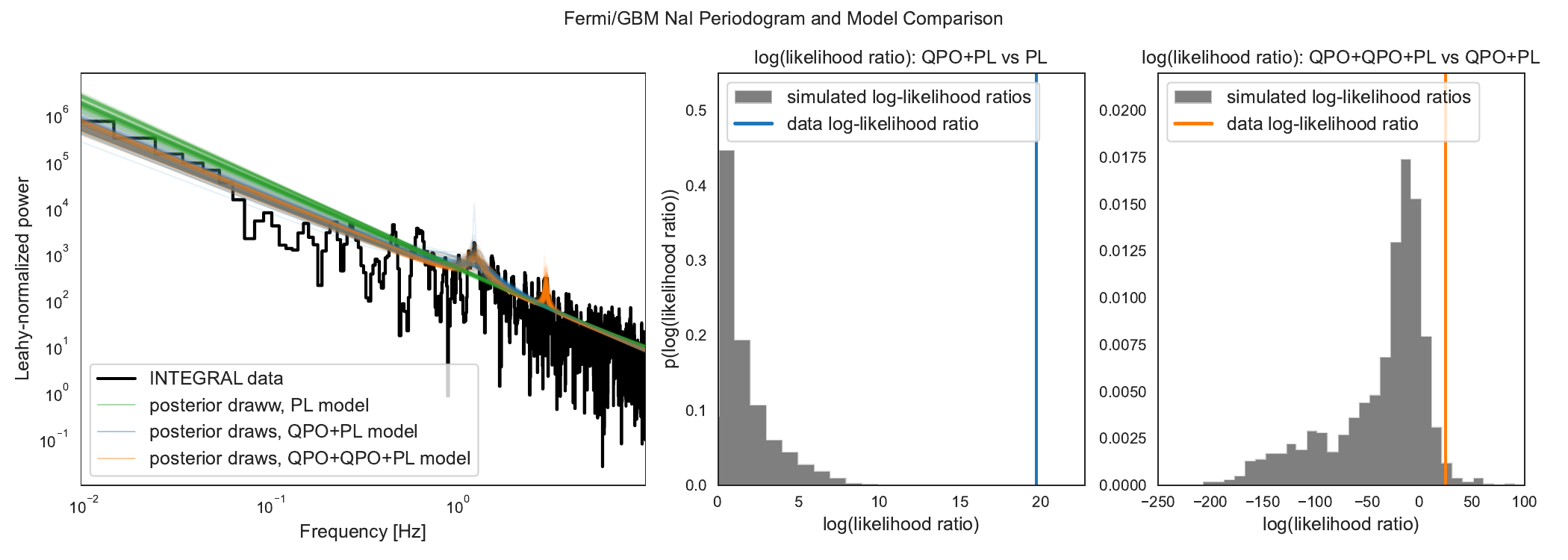}
    \caption{Left: Fourier periodogram of the \textit{Fermi}/GBM NaI data with posterior draws from the three models compared via likelihood ratio tests: in green, the power law model; in blue, a power law model with a Lorentzian component for a single QPO; in orange, a model comprising a power law and two Lorentzians. Middle: distribution of the likelihood ratios from 1000 simulated periodograms: the likelihood ratio for the observed periodogram is a clear outlier. Right: same as middle panel, but for the model with two QPOs. Again, the observed likelihood ratio is a clear outlier compared with the null hypothesis (a single QPO).}
    \label{fig:fermi_nai_periodogram}
\end{figure*}

\begin{figure*}[h!]
    \centering
    \includegraphics[width=\textwidth]{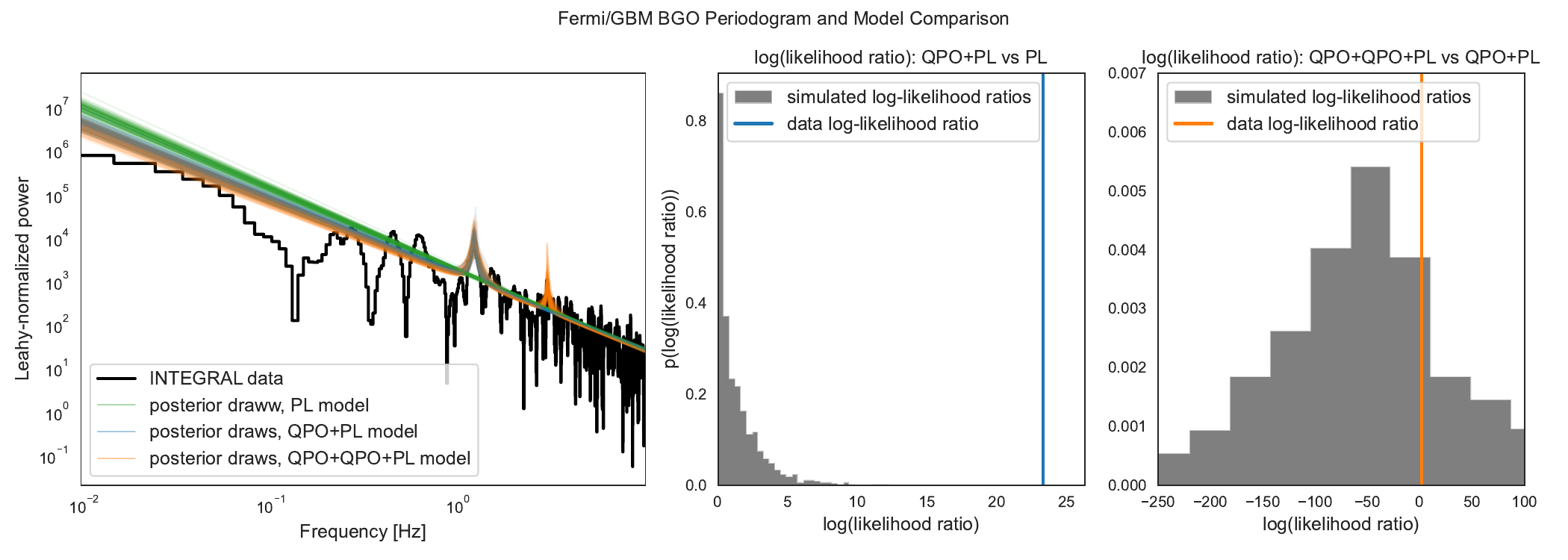}
    \caption{Left: Fourier periodogram of the \textit{Fermi}/GBM BGO data with posterior draws from the three models compared via likelihood ratio tests: in green, the power law model; in blue, a power law model with a Lorentzian component for a single QPO; in orange, a model comprising a power law and two Lorentzians. Middle: distribution of the likelihood ratios from 1000 simulated periodograms: the likelihood ratio for the observed periodogram is a clear outlier. Right: same as middle panel, but for the model with two QPOs. Again, the observed likelihood ratio is a clear outlier compared with the null hypothesis (a single QPO).}
    \label{fig:fermi_bgo_periodogram}
\end{figure*}

\begin{figure*}
    \centering
    \includegraphics[width=0.9\textwidth]{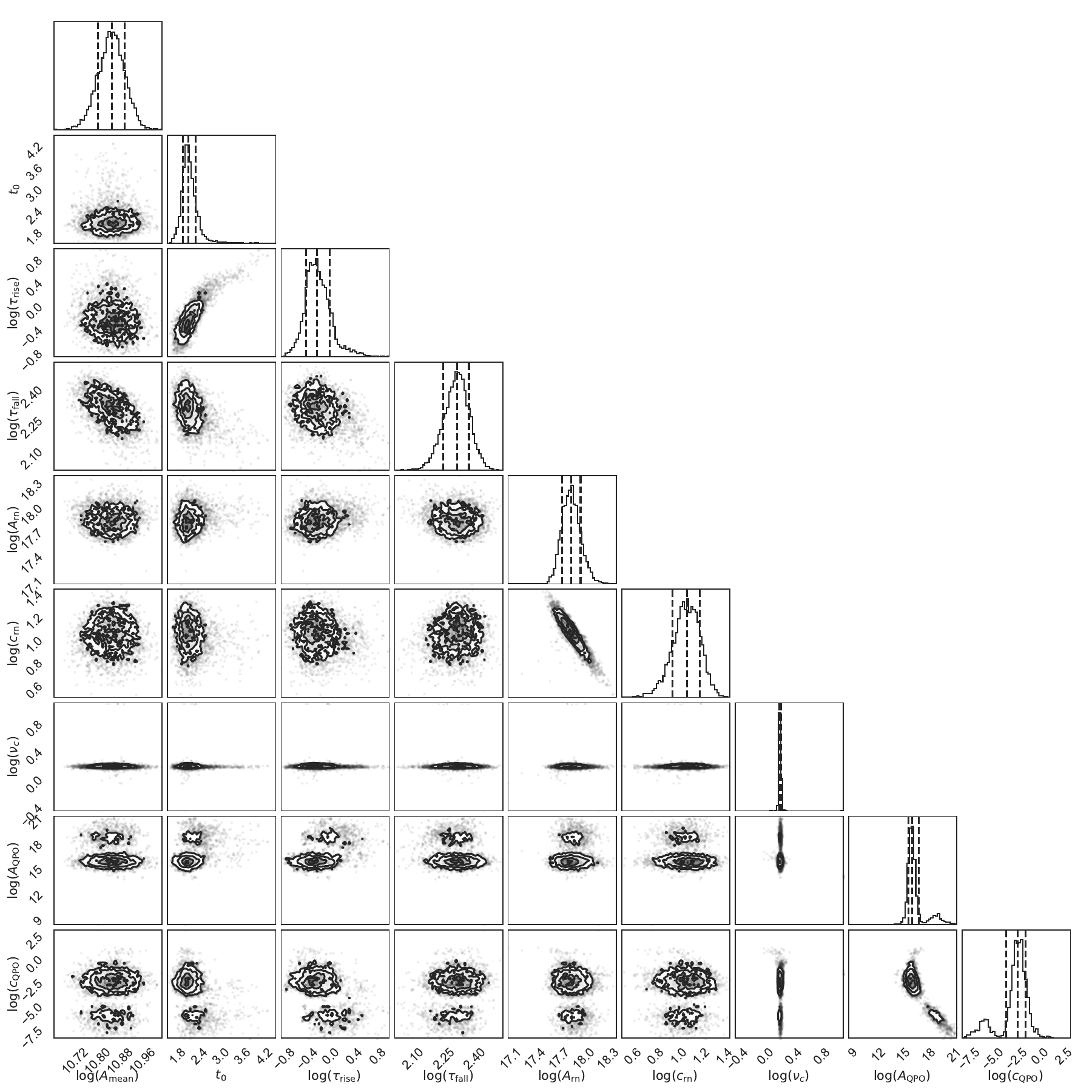}
    \caption{Posterior probability densities for all parameters in the DRW+QPO model for the SPI-ACS data described in Section \ref{sec:spi-acs}.}
    \label{fig:integral-drw-qpo-corner}
\end{figure*}

\begin{figure*}
    \centering
    \includegraphics[width=\textwidth]{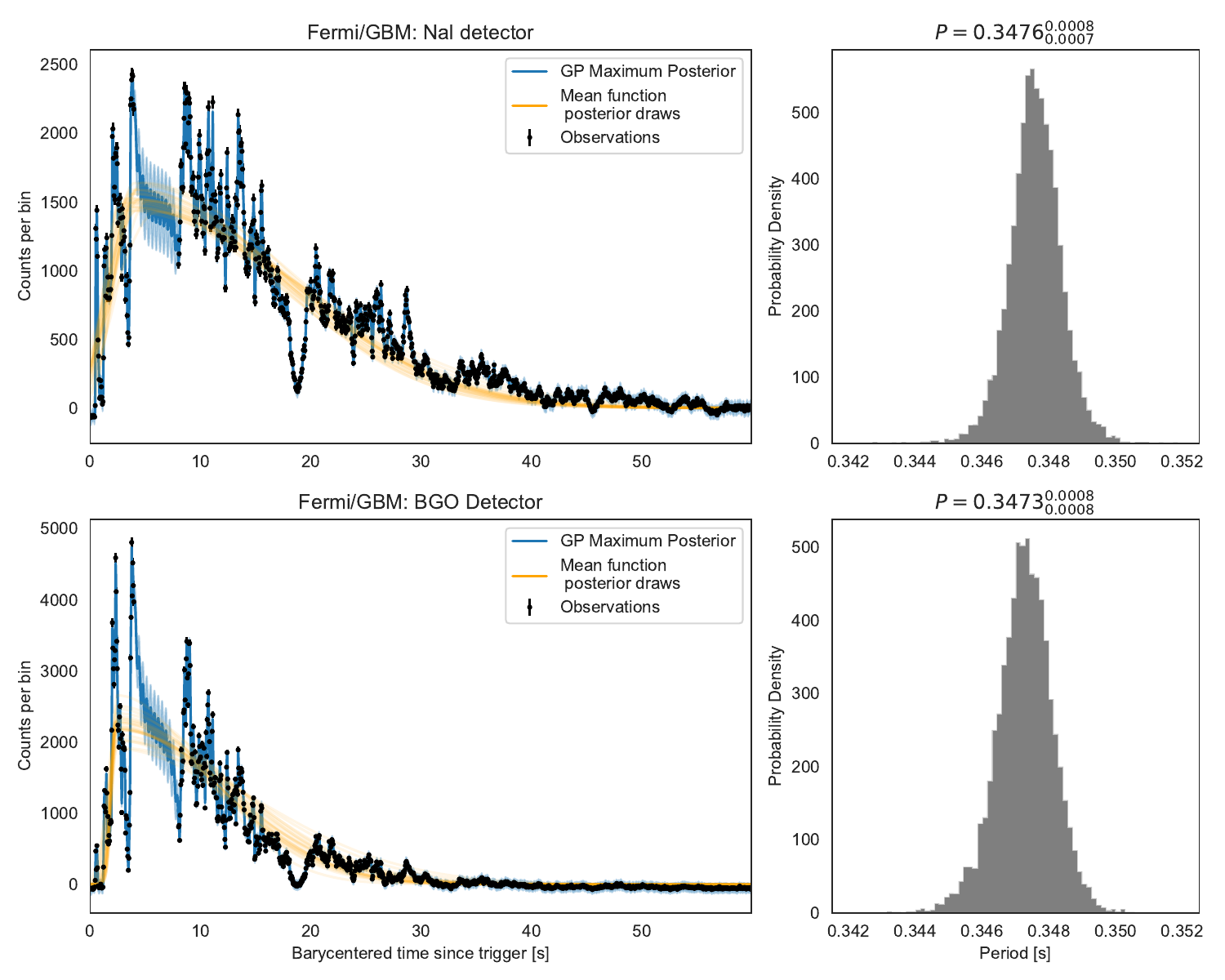}
    \caption{Left panels: light curve (black) with BTI removed, maximum-a-posteriori model (blue) and posterior draws from the mean model (orange) for the \textit{Fermi}/GBM NaI detector (na; top) and the BGO detector (b1; bottom). Right: corresponding posterior probability densities for the period in the model including a QPO. This figure suppements the results in Section \ref{sec:fermi_gp_qposearch}.}
    \label{fig:fermi-cutout-postpred}
\end{figure*}

\end{appendix}

\end{document}